\documentclass[twocolumn]{aastex7}

\usepackage{soul}
\usepackage{booktabs}
\usepackage[thinc]{esdiff}

\newcommand\der{\textrm{d}}

\begin{document}

\title{A Near-Infrared Extinction and Reddening Map Towards the Galactic Bulge Using UKIRT}

\author[0000-0002-6387-7729]{Aiden S. Zelakiewicz}
\affiliation{Department of Astronomy and Carl Sagan Institute, Cornell University, 122 Sciences Drive, Ithaca, NY 14853, USA}
\affiliation{Department of Astronomy, The Ohio State University, 140 West 18th Avenue, Columbus, OH 43210, USA}
\email{asz39@cornell.edu}

\author[0000-0001-9397-4768]{Samson A. Johnson}
\affiliation{Department of Astronomy, The Ohio State University, 140 West 18th Avenue, Columbus, OH 43210, USA}
\email{samson.a.johnson@gmail.com}

\author[0000-0003-0395-9869]{B. Scott Gaudi}
\affiliation{Department of Astronomy, The Ohio State University, 140 West 18th Avenue, Columbus, OH 43210, USA}
\email{gaudi.1@osu.edu}

\author[0000-0001-5966-837X]{Geoffrey Bryden}
\affiliation{Jet Propulsion Laboratory, 4800 Oak Grove Drive, Pasadena, CA 91109, USA}
\email{geoffrey.bryden@jpl.nasa.gov}

\author[0000-0001-5825-4431]{David M. Nataf}
\affiliation{Department of Physics \& Astronomy, University of Iowa, Iowa City, IA 52242, USA}
\email{david-nataf@uiowa.edu}

\author[0000-0003-1525-5041]{Yossi Shvartzvald}
\affiliation{Department of Particle Physics and Astrophysics, Weizmann Institute of Science, Rehovot 7610001, Israel}
\email{yossi.shvartzvald@weizmann.ac.il}

\begin{abstract}

The Galactic bulge is one of the most information-dense regions to study resolved stellar populations, variables, and transients, such as microlensing events.
Studies toward the Galactic bulge are complicated by the large and variable extinction along the line of sight.
We measure the near-infrared $A_{K_S}$ extinction and $E(H-K_S)$ reddening in this region using $H$- and $K$-band photometry obtained with the 2017 UKIRT microlensing survey. 
We fit the apparent magnitude and color distribution of bright giants in the bulge to recover the apparent magnitude and color of Red Clump stars, which are known to be standard candles and crayons.
We present $2^\prime \times 2^\prime$ resolution maps in UKIRT fields between $-2.15^{\circ} \le l \le 2.71^{\circ}$ and $-2.69^{\circ} \le b \le 2.03^{\circ}$ of the $A_{K_S}$ extinction and the $E(H-K_S)$ reddening.
We find large variations in the $K_S$-band extinction and $E(H-K_S)$ reddening on all the scales we probe.
We find that a constant, standard extinction law is a poor representation of the relationship between the extinction and reddening we measure in fields of different latitudes.
These maps will be useful for understanding the near-infrared extinction law for sight lines close to the Galactic plane, as well as for final field selection for the \textit{Nancy Grace Roman Space Telescope} Galactic Bulge Time Domain Survey.

\end{abstract}

\section{Introduction} \label{sec:intro}

The Galactic bulge is the most densely populated stellar region in the Milky Way (MW) and serves as a critical target for a variety of studies.
However, observations towards the bulge are plagued by large and highly variable amounts of interstellar extinction, caused by the absorption and scattering of light by dust along the line of sight.
Understanding the role and extent to which dust attenuates light sourced from the bulge is crucial for essentially all studies towards the bulge.

Specifically, a better handle on extinction has a positive domino effect on other areas.
Studies of the Galactic Center \citep{bovy2016, fritz2021, henshaw2023} and Sgr A* \citep{gillessen2009, genzel2010, john2024} require precise measurements of the interstellar extinction.
In addition, fields such as cosmology \citep{nataf2016, duarte2023}, microlensing \citep{fouque2010, yee2015, 2018ApJ...857L...8S, wen2023, koshimoto2023}, exoplanet detection \citep{sullivan2015, wilson2023}, and stellar characterization \citep{andrae2018, 2019MNRAS.490.3158C} depend heavily on accurate measurements of extinction and reddening.
Uncertainties in extinction and reddening measurements can propagate through these studies, leading to large uncertainties or even incorrect interpretations of a variety of phenomena.

Much work has been done to better map the extinction and reddening towards the bulge throughout the optical and infrared.
Large surveys have allowed the extinction and reddening to be mapped at high resolutions, tracing complex structures within the bulge \citep[e.g.,][]{green2019}.
One successful method of tracing extinction is using Red Clump (RC) stars as standard candles and crayons.
The Red Clump occupies a compact region of the color-magnitude diagram and, once the color and magnitude of the Red Clump are calibrated, can be used as a reference point of comparison to measure the impact of foreground extinction and reddening \citep{1996ApJ...464..233W, 1996ApJ...460L..37S, nataf2011, nataf2013}.

\citet{gonz2012} and \citet{wegg2013} utilized dual-band photometry from the Vista Variables in the Via Lactea (VVV) survey to measure the $(J-K_S)$ reddening in the bulge.
They compared the color of Red Clump stars in the bulge with RC stars in Baade's window, a well-studied region of known extinction, as standards \citep{baade1946, stanek1996}.
Building on this work, \citet{gonz2018} and \citet{surot2020} used the VVV with improved data reduction to also measure the reddening of Red Clump stars (further outlined in Sections \ref{sec:gonz} and \ref{sec:surot}, respectively).
These maps of color excess (we use ``reddening'' and ``color excess'' interchangeably in this work) are converted to extinction values using reddening laws such as that from \citet{ccm}.

While standard reddening laws such as \citet{ccm} are commonly used, they do not accurately represent the relationship between reddening and extinction for all lines of sight \citep[][]{2025Sci...387.1209Z}.
The conversion from reddening values to extinction is especially challenging in fields of high extinction, where reddening laws have been measured to be non-standard and vary with location.
While extinction laws can accurately describe moderately reddened fields in the MW, they deviate when applied to low and high-extinction populations.
Various independent studies such as \citet{zasowski2009}, \citet{nishiyama2009}, and \citet{nataf2016} have investigated this issue and all reported similar patterns in the mid-infrared, near-infrared (NIR), and optical, respectively.

Recent literature often opts to report only the color excess instead of extinction, moving the responsibility onto the user to select reddening laws.
A better alternative is to directly measure both the extinction and reddening, such as the work by \citet{nataf2013}.
They directly and semi-independently measure the $I$-band extinction and $(V-I)$ reddening using the OGLE-III gravitational lensing survey \citep{udalski2003b, udalski2008}.
They fit a magnitude distribution of a form similar to \citet{ps1998} to recover the mean $I$-band magnitude of the Red Clump as a function of position, allowing them to derive accurate maps of the NIR extinction and total-to-selective extinction.
Such maps will be a boon to upcoming time-domain surveys toward the bulge.

As NASA's next flagship astrophysics mission after \textit{JWST}, \textit{The Nancy Grace Roman Space Telescope} (\textit{Roman}; previously \textit{WFIRST}), will conduct several large Core Community Surveys, including a 438-day Galactic Bulge Time Domain Survey (GBTDS) of $\sim 1.4$ square degrees toward the Galactic bulge \citep{rotac}.
The GBTDS is expected to detect $\sim 27,000$ microlensing events \citep{2015arXiv150303757S,2019ApJS..241....3P} due to stars, stellar remnants \citep{lam2020}, bound planetary systems \citep{2019ApJS..241....3P} and free-floating planets \citep{2020AJ....160..123J, sumi2023}.
The survey will also detect $\sim 10^5$ transiting planets \citep{montet2017, wilson2023}, and will enable a broad range of additional science, such as the detection of transients and variable stars, astroseismology of bulge giants \citep{gould2015, Weiss:2025}, studies of Galactic stellar populations and structure, the detection of thousands of trans-Neptunian objects \citep{gould2014}, and more \citep[see, e.g.,][]{gaudi2019}.

While lines of sight toward the Galactic bulge have the highest microlensing event rates in our Galaxy \citep{kiraga1994, sumipenny2016} due to the density of lenses and sources, interstellar extinction towards these lines of sight is also extinguishing and reddening the light we observe.
This complicates the interpretation of microlensing light curves, because a measurement of the unextinguished flux and color of the source is necessary to break degeneracies in many observed microlensing events \citep{1999ApJ...512..672A, udalski2003, 2004ApJ...616.1204Y}.

\begin{figure}[t]
    \centering
    \includegraphics[width=\linewidth]{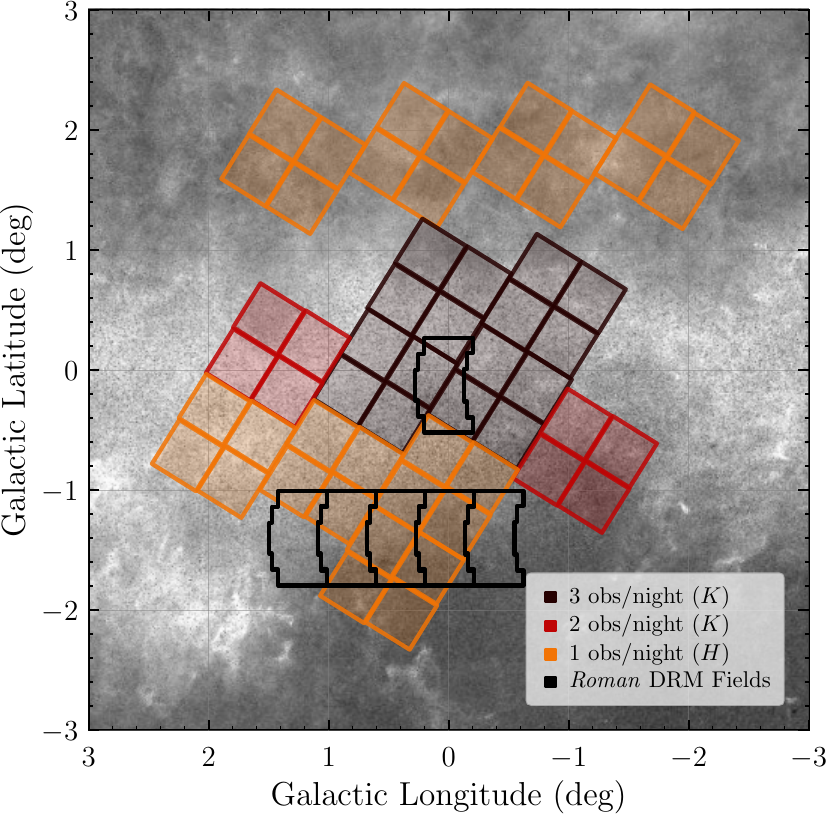}
    \caption{The fields observed by the UKIRT team during the 2017 and 2018 seasons. The 2019 season contains the same regions except for the top row of orange grids. Each square represents a grouping of the four CCDs on WFCAM. The black area shows the 438-day overguide scenario for the \textit{Roman} GBTDS fields \citep{2019ApJS..241....3P, rotac}. The background is the Milky Way as seen by the \textit{Gaia} Data Release 3 using \texttt{mw-plot} \citep{mw-plot}.}
    \label{fig:ukirtloc}
\end{figure}

The United Kingdom Infrared Telescope (UKIRT) performed a pathfinder microlensing survey for \textit{Roman} from 2015 to 2019 \citep{2017AJ....153...61S}.
The UKIRT microlensing survey's objective was to measure the microlensing event rate in the NIR \citep{wen2023}.
Even though interstellar dust has a lower impact on the NIR than the optical, it still extinguishes significant portions of light across the bulge.
Understanding the extent of extinction's impact is crucial due to the impact extinction has on being able to measure the microlensing event rate and characterize individual events \citep[][]{gaudi2012, yee2015}.

In this paper, we use archival data from the UKIRT microlensing survey to map the $K_S$-band extinction and $E(H-K_S)$ reddening towards the bulge.
We describe the data we used to fit for the magnitude and color and our quality control process in Section \ref{sec:data}.
In Section \ref{sec:fit} we overview the methodology we used, following \citet{Nataf2013Fit, nataf2013}, to isolate the Red Clump and measure its $K_S$-band magnitude and $(H-K_S)$ for each line of sight.
In Section \ref{sec:result}, we discuss our results in the context of various NIR extinction and reddening maps.
The validity of a universal extinction law on our dataset and map is also explored here.

\section{Data} \label{sec:data}

UKIRT photometry data was obtained at the UKIRT 3.8m telescope on Mauna Kea, Hawaii.
Observations employed the near-infrared Wide Field Camera (WFCAM), which comprises four detectors, each covering a $13.6' \times 13.6'$ field of view \citep{2017AJ....153...61S}.
Figure~\ref{fig:ukirtloc} illustrates the UKIRT fields surveyed during the 2017-2019 observing seasons.
We analyze $H$‑ and $K$‑band data from 2017 in this work.

Field cadence is indicated by color in Figure~\ref{fig:ukirtloc}: light orange fields were observed once per night in $H$‑band and once every five nights in $K$‑band; red fields twice per night in $K$‑band and once every three nights in $H$‑band; and dark crimson fields three times per night in $K$‑band and once every three nights in $H$‑band \citep{2017AJ....153...61S}.
The surveys were able to obtain observations in the magnitude range of 11.5 to 18 mag for the $K$-band and 11.5 to 19 mag for the $H$-band\footnote{UKIRT magnitude limits: \url{https://exoplanetarchive.ipac.caltech.edu/docs/UKIRT_figures.html}}.
This rough approximation is field-dependent due to highly variable differential extinction in the galactic plane.
For further details on the telescope and instrumentation, see \citet{2007MNRAS.379.1599L}.

We derive extinction and reddening by constructing color-magnitude diagrams of stars within square regions of a specified edge length, which also define the pixel scale of our maps (see Section~\ref{sec:map}).
We select all stars whose right ascension and declination lie inside each region.
Photometric light curves were obtained via two methods: aperture photometry from the CASU 2MASS-calibrated catalog \citep{2009MNRAS.394..675H}, and PSF photometry performed with SExtractor and PSFEx \citep{1996A&AS..117..393B, bertin2011}, also calibrated to 2MASS.
In this work, we adopt the PSF photometry because it is more robust than the CASU photometric catalogs for measuring faint objects in crowded fields \citep{2009MNRAS.394..675H,2017AJ....153...61S, 2018ApJ...857L...8S}.

\begin{figure}[t]
    \centering
    \includegraphics[width=\linewidth]{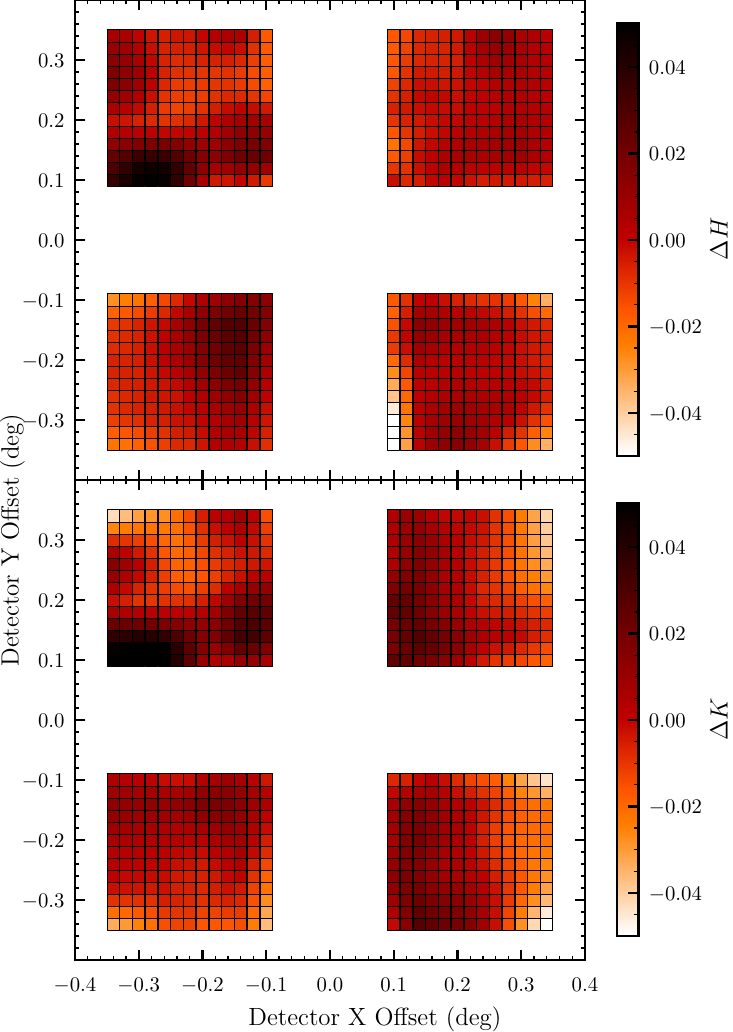}
    \caption{Illumination corrections for the 2017 UKIRT microlensing survey using the May and July 2017 products.}
    \label{fig:illum}
\end{figure}

Spatial and temporal systematics are present in WFCAM observations that require a detector-level calibration.
The UKIRT microlensing survey was conducted between April and August 2017, where corrections supplied by CASU for May and July of that year are available \citep{2009MNRAS.394..675H}.
Figure~\ref{fig:illum} is the average of the July and May corrections for the $H$- and $K$-band.
We apply the averaged detector corrections to the $H$- and $K$-band observations using bilinear interpolation based on each star's location in the CCD.

\begin{figure}
    \centering
    \includegraphics[width=\linewidth]{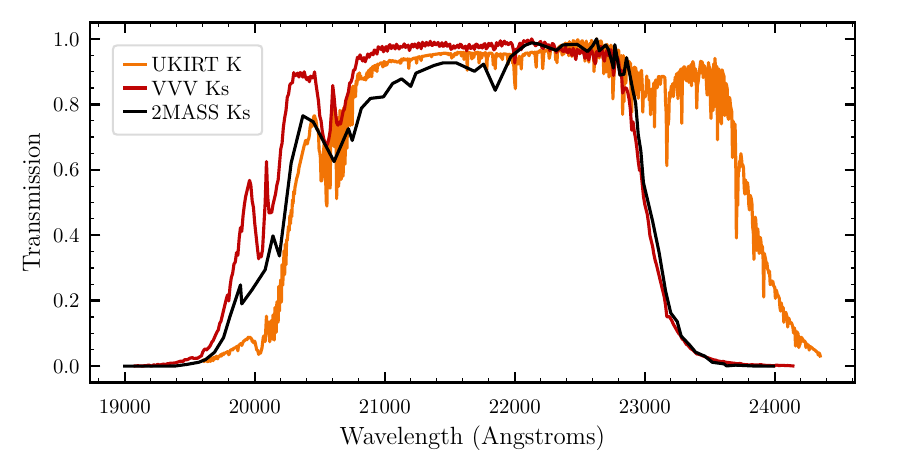}
    \caption{$K$-band filter used by the UKIRT microlensing survey as well as $K_S$-band filters from VVV and 2MASS with transmission normalized to $1.0$ for all filters. Filter profiles were provided by the Spanish Visual Observatory's (SVO) Filter Profile Service \citep{2020sea..confE.182R}.}
    \label{fig:filter}
\end{figure}

We impose quality control constraints to filter out poorly characterized stars.
All stars must have $\geq10$~$K$-band and $\geq3$~$H$-band measurements, and the median average dispersion in $H$- and $K$-band magnitudes must be $\leq0.1$ mag to avoid stars afflicted by systematic errors in photometric extraction.
We also require that the $H$- and $K$-band measurements agree to within $0.1^{\prime\prime}$ in their right ascension and declination.

\begin{figure}[t]
    \centering
    \includegraphics[width=\linewidth]{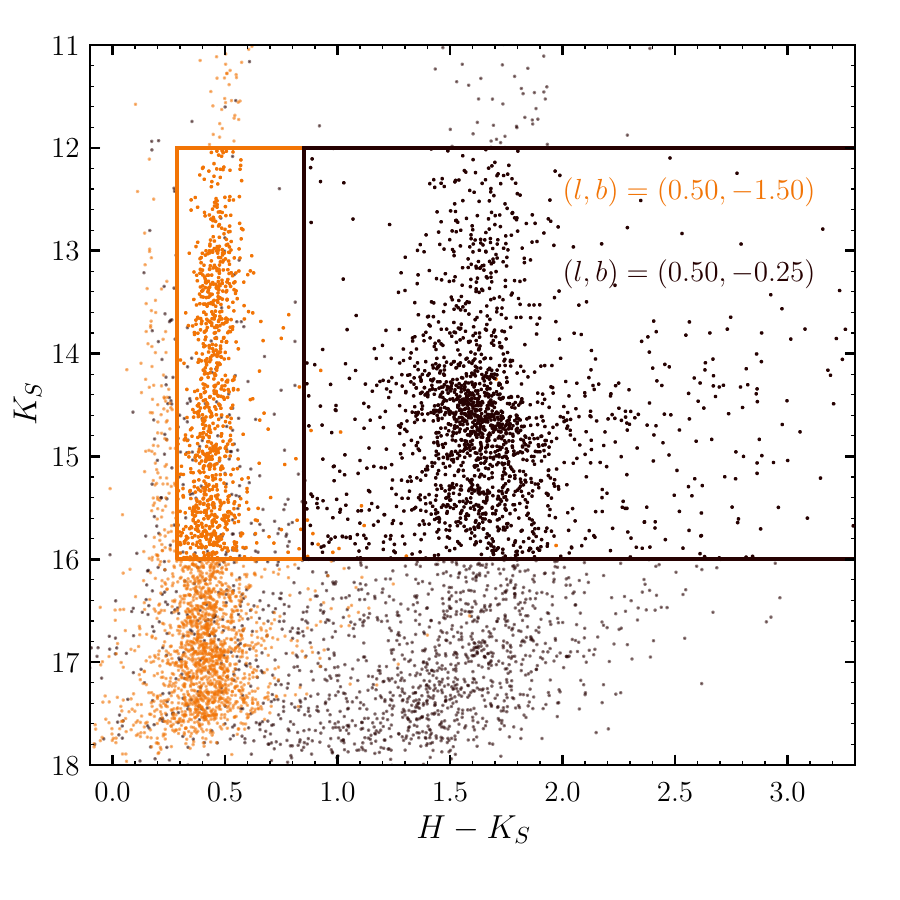}
    \caption{
    Shown are two example fields of stars that passed the quality cuts.
    The brown is a field close to the GC with significant extinction and reddening, whereas the orange is a field away from the Galactic Plane where extinction and reddening have a lesser effect.
    The boxes enclose stars used in fitting the luminosity function to isolate the Red Clump outlined in Section \ref{sec:fit}.
    }
    \label{fig:CMDExample}
\end{figure}

The CASU calibration process applied additional 2MASS color corrections compared to the PSF system, making the calibration more robust.
We compare the PSF and CASU catalogs using bright stars (12-14 mag) that pass the quality control cuts to align the PSF system to CASU.
We find $\tilde{H}_\mathrm{CASU} = \tilde{H}_\mathrm{PSF} + 0.025$ mag and $\tilde{K}_\mathrm{CASU} = \tilde{K}_\mathrm{PSF} - 0.029$ mag and apply the correction to the PSF photometry we adopt.
All the stars in the example color-magnitude diagrams in Figures~\ref{fig:CMDExample} and \ref{fig:6cmd} satisfy these criteria.

The $K_S$-band filters used by both VVV and 2MASS, to which the CASU products were calibrated, are bluer than the $K$-band filter of WFCAM on UKIRT, shown in Figure~\ref{fig:filter}.
As all of our photometry is ultimately calibrated to 2MASS, we quote our results (extinction maps, etc.) in the 2MASS $K_S$ filter system, despite the fact that the data were, in fact, taken in the UKIRT WFCAM $K$ filter. 
Since the CASU calibration of the UKIRT WFCAM $K$ to 2MASS $K_S$ included color terms, this difference is properly accounted for.

\section{Fitting for Extinction and Reddening} \label{sec:fit}

Here, we describe the methods we use to measure the extinction and reddening of stars within fields of the UKIRT survey footprint.
To collect the most Red Clump stars for the fit, we include stars within a $K_S$-band magnitude range of [12,16] and also stars that are above the 15th percentile in $(H-K_S)$ color.
Two example fields, one high and one low latitude, are given in Figure~\ref{fig:CMDExample} with the box enclosing stars used in the final fit.

\subsection{Luminosity Function} \label{sec:lum}

The luminosity function given in Equation \ref{eqn:lum_func} describes the number of stars $N(K_S)$ per magnitude $K_S$ \citep{Nataf2013Fit, nataf2013}.
\begin{eqnarray}
\label{eqn:lum_func}
\diff{N(K_S)}{K_S} &=& A\exp\left[B(K_S-K_{S,RC})\right]\nonumber\\
&+&\frac{N_{RC}}{\sqrt{2 \pi}\sigma_{RC}} \exp \left[-\frac{(K_S-K_{S,RC})^2}{2\sigma_{RC}^2}\right]\nonumber\\
&+&\frac{N_{RGBB}}{\sqrt{2\pi}\sigma_{RGBB}} \exp \left[-\frac{(K_S-K_{S,RGBB})^2}{2\sigma_{RGBB}^2}\right] \nonumber\\
&+&\frac{N_{AGBB}}{\sqrt{2\pi}\sigma_{AGBB}} \exp \left[-\frac{(K_S-K_{S,AGBB})^2}{2\sigma_{AGBB}^2}\right]
\end{eqnarray}

$N(K_S)$ is a function of the average magnitude of the Red Clump in the $K_S$-band $K_{S,RC}$, the total number of Red Clump stars $N_{RC}$ and its dispersion $\sigma_{RC}$.
The exponential in Equation \ref{eqn:lum_func} represents the background bulge stars, with $B$ parameterizing the gradient of their distribution in $K_S$ and $A$ setting the normalization \citep{castellani1989, Nataf2013Fit, nataf2013}.
There are three Gaussian terms included, the first of which describes the distribution of Red Clump stars.
The second and third Gaussian terms are for the Red Giant Branch Bump (RGBB) and Asymptotic Giant Branch Bump (AGBB), respectively.

\subsection{Reduction of Fit Parameters} \label{sec:params}

To increase the robustness of the fitted parameters of interest and decrease the computational cost of fitting, we reduce the number of parameters in Equation~\ref{eqn:lum_func} by defining a relationship between the RGBB and AGBB with the Red Clump.
We relate the RGBB and AGBB components to those of the Red Clump by imposing:
\begin{eqnarray}
N_{RGBB}&=&\delta_{RGBB} N_{RC} \nonumber\\
N_{AGBB}&=&\delta_{AGBB} N_{RC} \nonumber\\
K_{S,RGBB}&=&\Delta K_{S,RGBB} + K_{S,RC} \label{subeqn:rgbbK}\\
K_{S,AGBB}&=&\Delta K_{S,AGBB} + K_{S,RC} \nonumber\\
\sigma_{RGBB} &=& \sigma_{AGBB} = \sigma_{RC} \nonumber
\end{eqnarray}
The RGBB and AGBB are secondary features in our luminosity profiles when compared to the Red Clump and background bulge, and we do not aim to probe their individual characteristics.

\begin{figure*}[t]
    \centering 
    \gridline{\fig{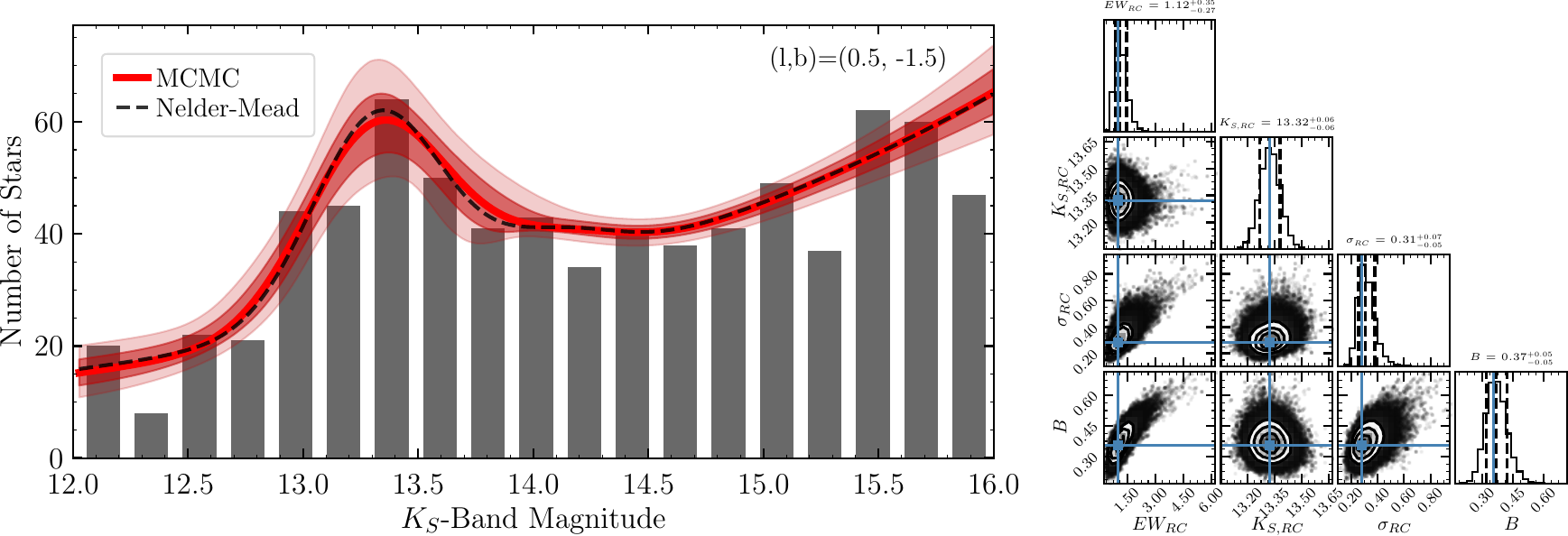}{\linewidth}{(a)}}
    \gridline{\fig{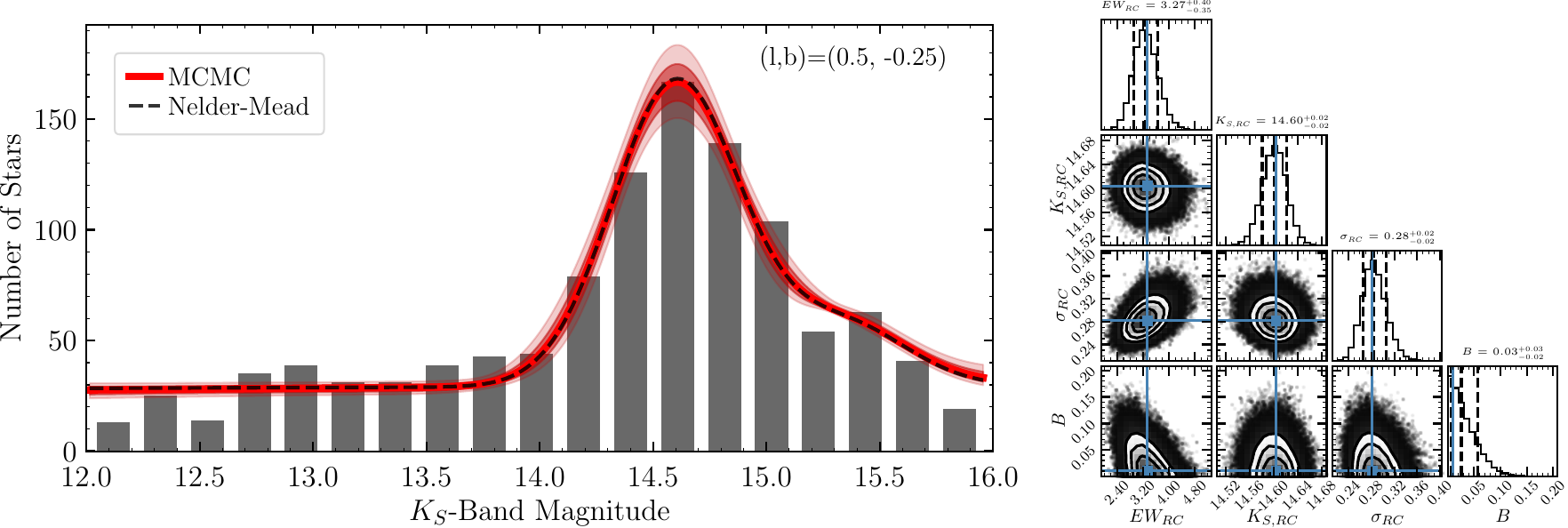}{\linewidth}{(b)}}
    \caption{Fit results for two sample fields of standard, $2^\prime \times 2^\prime$ size. We stress that we do not bin the data in our analysis and that binning is solely for visualization. Panel (a) is a high-latitude field that effectively characterizes the Red Clump and background giant stars. Panel (b) is a low-latitude field that suffers from crowding and completeness issues. The exponential shape of the background giant branch stars is missing, resulting in $B$ pushing against the prior lower limit to account for the lack of stars. This field is an example where we only report a lower limit on the extinction. Corner plots of the MCMC sampling are provided in the right panels, with blue lines indicating values from the Maximum Likelihood Estimation.}
    \label{fig:fits}
\end{figure*}

We assume values for $\delta_{RGBB}$, $\Delta K_{S, RGBB}$, $\delta_{AGBB}$, $\Delta K_{S, AGBB}$ are comparable to those for the $I$-band and adopt them within our work (0.201, 0.737, 0.028, and -1.07, respectively; \citealt{nataf2013}).
We tested these assumptions by constructing $30^\prime \times 30^\prime$ color-magnitude diagrams in multiple lines of sight.
With significantly more stars in these test fields, we find the relative location and number of RGBB stars with respect to the Red Clump in $K$-band to be consistent with those in the $I$-band.
We also do not find significant contributions from the AGBB in our color-magnitude diagrams at any scale, shown in the example fields in Figure~\ref{fig:fits}.
Due to the weak presence of the RGBB and AGBB, the adopted values for $\delta_{RGBB/AGBB}$ and $\Delta K_{S, RGBB/AGBB}$ won't significantly affect the fitted values for the $K_{S, RC}$.
This allows our luminosity function to only depend on variables associated with the Red Clump and exponential background: $A$, $B$, $K_{S,RC}$, $N_{RC}$, and $\sigma_{RC}$.

We rewrite Eq. \ref{eqn:lum_func}, following \citet{Nataf2013Fit}, as $\der {N(K_S)}/\der{K_S} = A (\der{N_{norm}(K_S)}/\der{K_S})$ where we define
\begin{eqnarray}
    \diff{N_{norm}(K_S)}{K_S}
    &\equiv& \exp\left[B(K_S-K_{S,RC})\right]\\
&+&\frac{EW_{RC}}{\sqrt{2 \pi}\sigma_{RC}} \exp\left[-\frac{(K_S-K_{S,RC})^2}{2\sigma_{RC}^2}\right]\nonumber\\
&+&\frac{EW_{RGBB}}{\sqrt{2\pi}\sigma_{RGBB}} \exp \left[-\frac{(K_S-K_{S,RGBB})^2}{2\sigma_{RGBB}^2}\right] \nonumber \\
&+&\frac{EW_{AGBB}}{\sqrt{2\pi}\sigma_{AGBB}} \exp \left[-\frac{(K_S-K_{S,AGBB})^2}{2\sigma_{AGBB}^2}\right] \nonumber
\end{eqnarray}
and the ``equivalent width'' of the Red Clump feature to be $EW_{RC}\equiv N_{RC}/A$, which leads to $EW_{RGBB} \equiv N_{RGBB}/A = \delta_{RGBB} EW_{RC}$ and similarly $EW_{AGBB} \equiv \delta_{AGBB} EW_{RC}$.
We then demand that
\begin{equation}
N_{obs} = N_{exp} = A\int \diff{N_{norm}(K_S)}{K_S} \der K_S
\end{equation}
which requires the integral of Eq. \ref{eqn:lum_func} over the range of $K_S$ we fit is equal to the total number of expected giant stars, $N_{exp}$, in this range of $K_S$ which we force to be equal to the number of stars included in the fit, $N_{obs}$.
This then sets $A$ for any value of $EW_{RC}$, $B$, $K_{S,RC}$ and $\sigma_{RC}$.
Thus we have only these four fit parameters remaining.

\subsection{Maximum Likelihood Estimation} \label{sec:minimize}

Stars that pass the quality control cuts set in Section \ref{sec:data} are used to determine the parameters in the luminosity function.
Following \citet{Nataf2013Fit}, we aim to optimize the luminosity function parameters utilizing the following log-likelihood function:
\begin{eqnarray} \label{eqn:likelihood}
    \mathcal{L} = \sum_i^{N_{obs}} \ln [ \der N(K_{S,i})/ \der K_S ] - N_{obs}
\end{eqnarray}
We optimize this log-likelihood in two steps: Maximum Likelihood Estimation (MLE) of Eq. \ref{eqn:likelihood} using Nelder-Mead \citep{nelder-mead} and exploring the parameter space using a Markov Chain Monte Carlo (MCMC).

We use an iterative process for the MLE to converge on the best parameter values.
We utilize the Python package \texttt{lmfit} \citep{lmfit}, a wrapper of \texttt{SciPy} \citep{scipy}, to conduct the Nelder-Mead optimization.
We perform an optimization using a fixed magnitude range of [12, 16], as described in Section \ref{sec:fit}.
The lower magnitude limit allows brighter stars in the exponential tail to contribute to the MLE and is much brighter than the expected $K_{S,RC}$ in the absence of extinction.
When the range is too tight, brighter stars get cut off, occasionally resulting in the optimizer failing to detect the background exponential.
The upper magnitude limit boundaries where the UKIRT survey suffers from incompleteness.
This effect is most notable in fields with high extinction, as can be seen in the lower panel of Figure~\ref{fig:fits} and further discussed in Section~\ref{sec:complete}.
We also use a universal initial guess for the MLE optimization, as outlined in Table~\ref{tab:mcmc}.

The recovered value for the $K_{S,RC}$ is compared against the seed value input into the MLE.
The MLE is run again if these values differ by more than 0.01 mag.
Initial conditions for this next iteration are set to be the results of the previous Nelder-Mead optimization.
This process is repeated until this magnitude convergence condition is met and rarely takes more than one iteration.

\subsection{Markov Chain Monte Carlo} \label{sec:mcmc}

\begin{table}[t]
\caption{Initial conditions for the initial MLE estimation and priors for each parameter in the MCMC run.}\label{tab:mcmc}
\begin{tabular}{@{}lcc@{}}
\toprule
Variable       & MLE    & MCMC Prior            \\ \midrule
$EW_{RC}$      & 2.0    & $\mathcal{U}(0, 10)$  \\
$B$            & 0.43   & $\mathcal{U}(0, 1)$   \\
$K_{S,RC}$       & 14.0   & $\mathcal{U}(12, 16)$ \\
$\sigma_{RC}$  & 0.5    & $\mathcal{U}(0, 1)$   \\ 
\bottomrule
\end{tabular}
\end{table}

Next, we explore the posterior distributions of the parameter space using the MCMC Python package, \texttt{emcee} \citep{emcee}.
Our MCMC runs were performed using $50$ walkers, $100,000$ iterations, and a burn-in of 1,000 iterations with initial conditions sourced from the converged MLE results.
The priors we set on $EW_{RC}$, $B$, $K_{S,RC}$, and $\sigma_{RC}$ are given in Table \ref{tab:mcmc}.
We chose wide uniform priors on our set of parameters, based on distributions from initial MLE results and theoretical limits.
Namely, the background exponential must be positive and increasing, bounding the lower limit on $EW_{RC}$ and $B$ \citep{castellani1989, Nataf2013Fit, nataf2013}.
We set a convergence criterion that the chain is 100 times longer than the autocorrelation time and that the autocorrelation time does not change by more than $1\%$ between iterations.

Figure~\ref{fig:fits} shows the best-fit luminosity functions from the MLE (black dashed) and the MCMC exploration (red) for the two sample fields in Figure~\ref{fig:CMDExample}.
The parameter distributions from the MCMC sampling are shown to the right of both fits, with the input MLE values overlaid as blue lines.
The average difference in best-fit values between the MLE and the MCMC-derived posterior medians across all fits is on the order of $10^{-3}$ magnitudes.
We therefore adopt the MCMC results for the reported values in our map.

Specifically, we report the 50th percentiles (medians) of the post-burn-in marginalized posterior distributions for each parameter.
We also report uncertainties as the deviations between the 16th and 50th percentiles and the 84th and 50th percentiles.
However, we note that the marginal medians do not necessarily correspond to the best fit to the data when considered jointly, since they are derived from one-dimensional marginal distributions.
As noted by \citet{hogg2018}, even for well-behaved posteriors, the collection of marginal medians may not correspond to a point of high posterior probability in the full parameter space.
For this reason, we emphasize that these reported values should be interpreted as informative summaries of individual parameters, not as a global optimum.

\begin{figure}
    \centering
    \includegraphics[width=\linewidth]{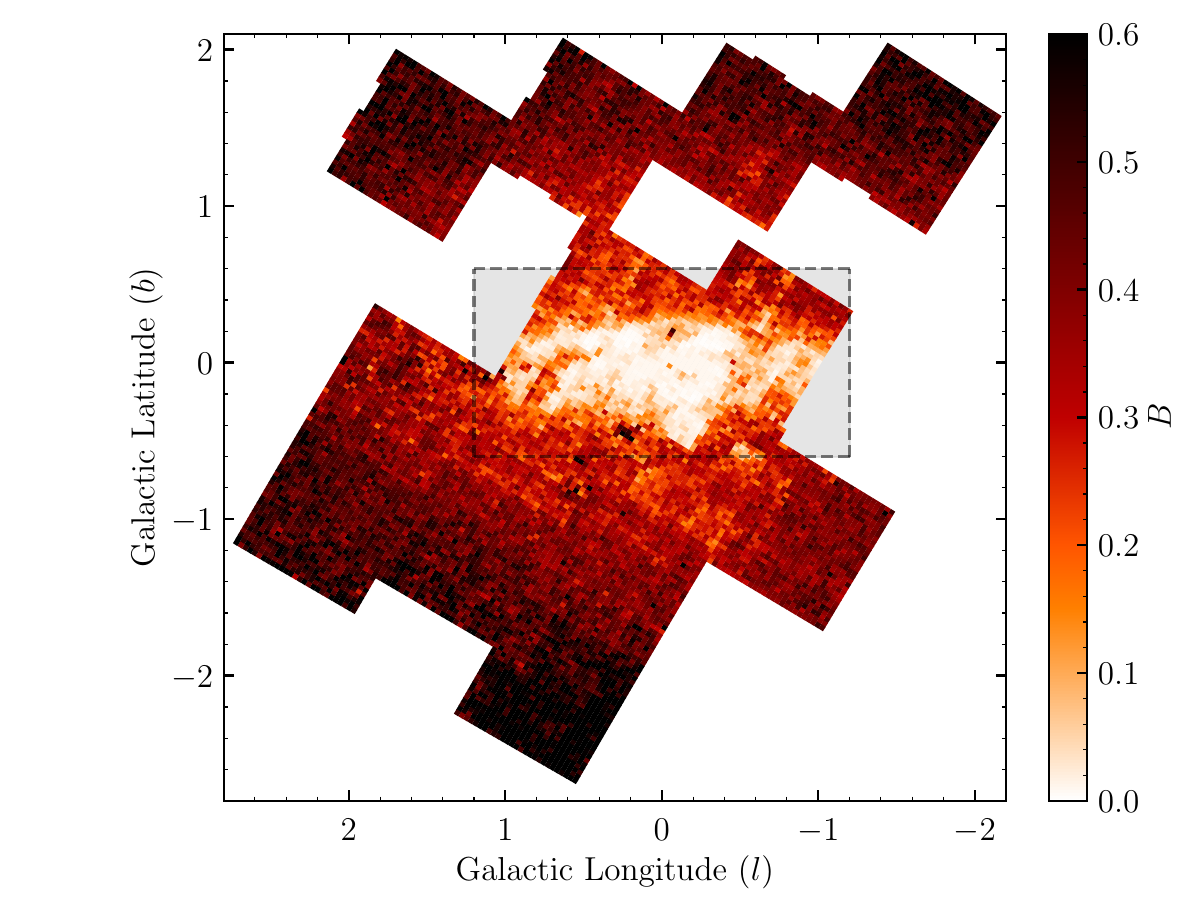}
    \caption{Mapping of the exponential gradient parameter, $B$, from our fits in UKIRT fields. We advise caution when using pixels with $B < 0.2$.}
    \label{fig:bmap}
\end{figure}

\begin{figure*}
    \centering
    \includegraphics[width=\linewidth]{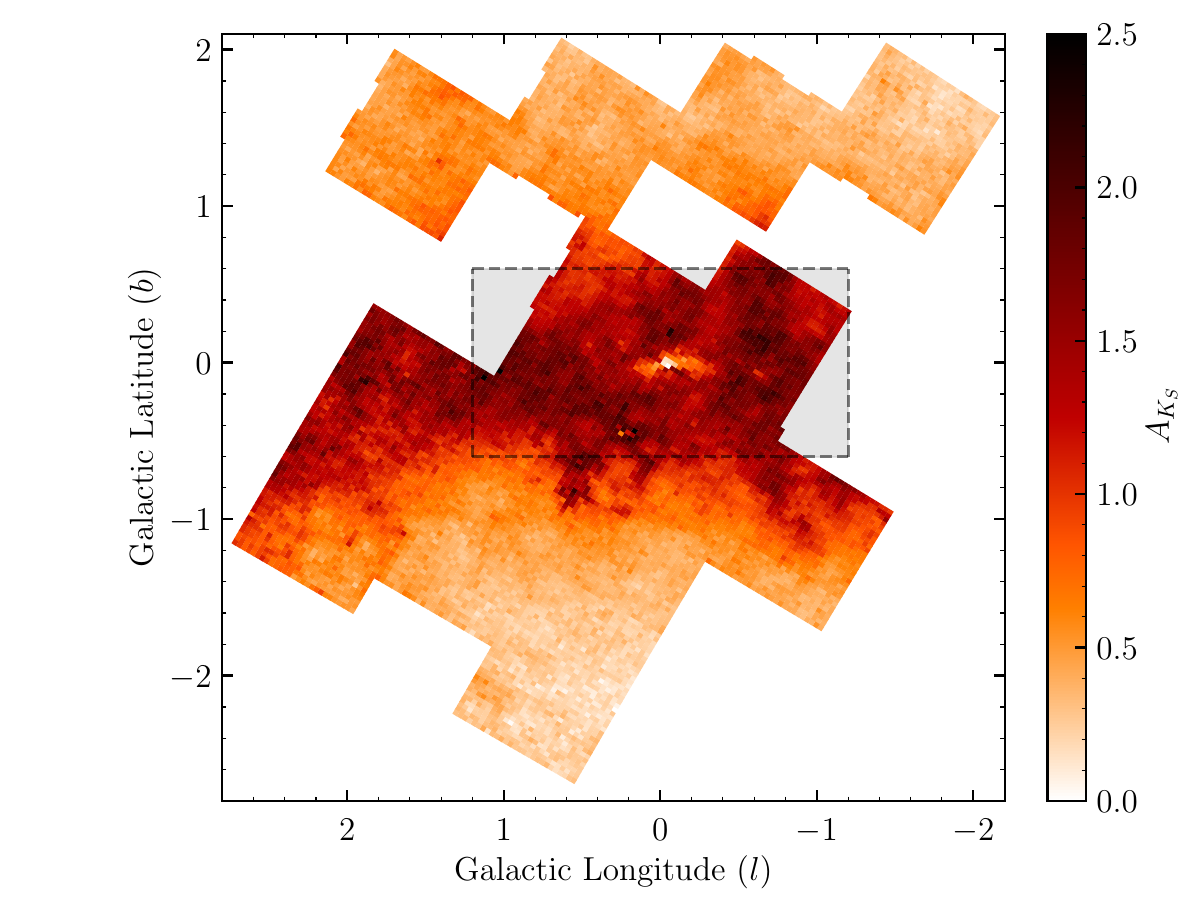}
    \caption{Map of the $A_{K_S}$ extinction of the Red Clump within the UKIRT field. This map uses the 2017 PSF photometry data with a $2^\prime$ edge length. Each pixel on the map is therefore of size $2^\prime \times 2^\prime$. The grayed region within $|b|\leq 0.6^{\circ}$ and $|l|\leq1.2^{\circ}$ is where we only declare lower limits on the extinction. The gray shading sits behind the map and does not affect the color of the pixels themselves.}
    \label{fig:finalAK}
\end{figure*}

\subsection{$(H-K_S)$ Color} \label{sec:color}

We utilize a brute force method to determine the color of the Red Clump for a given color-magnitude diagram following \citet{nataf2013}.
We begin by finding the color where the density of bulge stars is a maximum, which we initialize to be the guess for the Red Clump color $(H-K_S)_{RC,obs}$.
We also create a set of trial colors for the foreground disk between $(H-K_S)_{RC,obs} -2.5 \leq (H-K_S)_{\mathrm{disk}}$ and $(H-K_S)_{\mathrm{disk}}<(H-K_S)_{RC,obs} - 0.01$ with a spacing of 0.01 mag.
The lower bound on this set is also constrained by the requirement that the color is strictly greater than zero.
The difference between each star's color and both the Red Clump color and the trial disk color are computed.
Each star is then assigned to either the disk or clump based on which population its color more closely matches.
We assign weights to each star in the Red Clump group
\begin{eqnarray}
W_i = \frac{\frac{N_{RC}}{\sqrt{2 \pi}\sigma_{RC}} \exp\left[-\frac{(K_{S,i}-K_{S,RC})^2}{2\sigma_{RC}^2}\right]}{N(K_{S,i})/dK_{S,i}}
\end{eqnarray}
which compares the Red Clump Gaussian term of Eq. \ref{eqn:lum_func} to the full luminosity function.
We calculate the weighted average color of the Red Clump group $(H-K_S)_{RC,obs}$ and find the weighted variance of each star compared to $(H-K_S)_{RC,obs}$.
Any stars that are $2.5\sigma$ away from $(H-K_S)_{RC,obs}$ are removed from the sample and the process is repeated iteratively for each trial color until no more stars remain outside the $2.5\sigma$ range.
We then finalize a value of $(H-K_S)_{RC,obs}$ for the specific run that minimizes the weighted variance across all the trial foreground disk colors.
This process is repeated, where the accepted value of $(H-K_S)_{RC,obs}$ becomes the initial Red Clump color reference, until the initial color and variance-minimized color agree to within 0.01 mag.
If it takes more than 5 iterations to converge on a color, we flag the field and do not include its color in our final analysis.

\subsection{Completeness}\label{sec:complete}

The completeness for the UKIRT microlensing survey as a function of source magnitude has not been measured. 
UKIRT has an increasing incompleteness for higher magnitude stars and, combined with high amounts of extinction, does not detect significant fractions of the Red Clump, Red Giant Branch Bump, and background bulge stars in fields close to the Galactic Plane. 
The lack of stars in each of these groups has different but compounding effects on our determination of the extinction and reddening.

\begin{figure*}[ht]
    \centering
    \includegraphics[width=\linewidth]{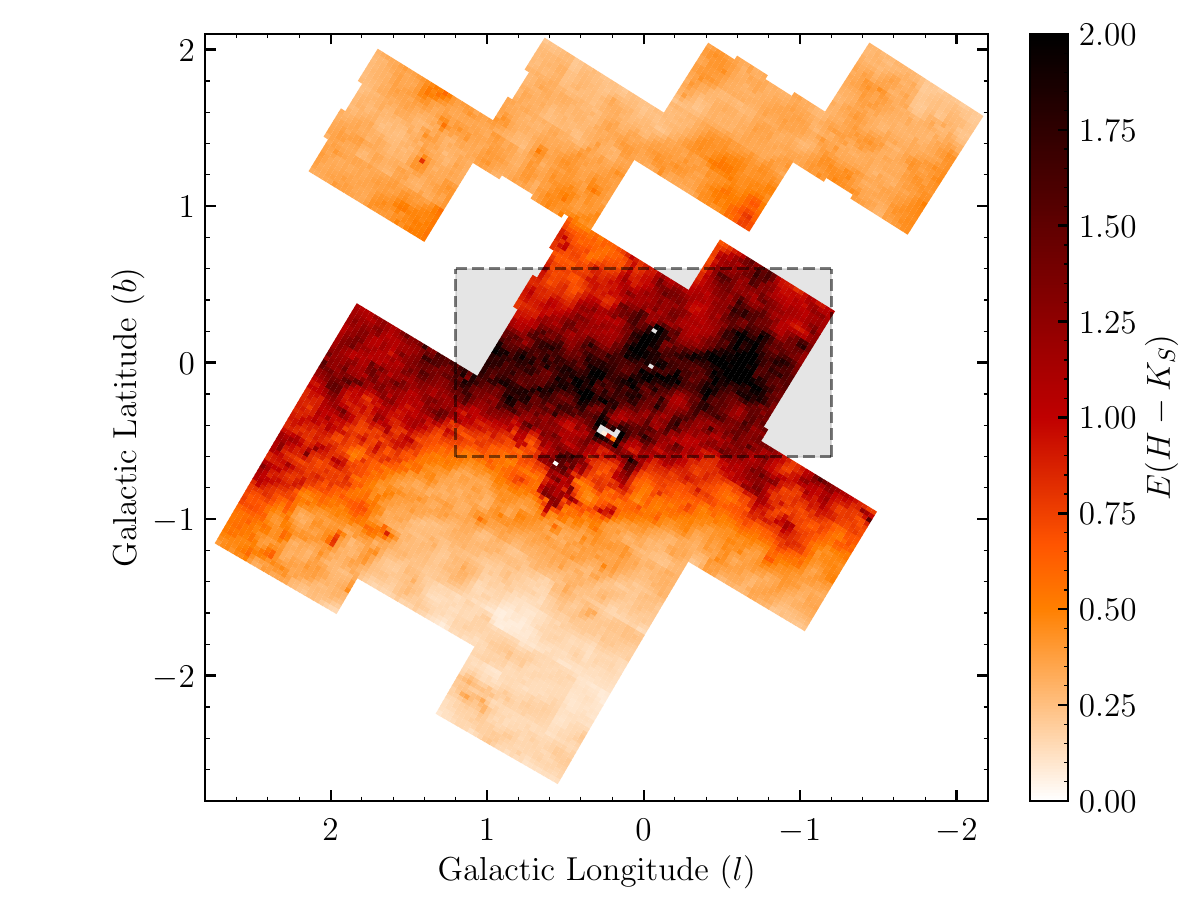}
    \caption{$2^\prime \times 2^\prime$ map of the $E(H-K_S)$ reddening of the Red Clump using $H$- and $K_S$-band UKIRT photometry. The grayed region within $|b|\leq 0.6^{\circ}$ and $|l|\leq1.2^{\circ}$ denotes where we declare lower limits on the extinction. The gray shading sits behind the map and does not affect the color of the pixels.}
    \label{fig:finalEHK}
\end{figure*}

The loss of faint background giant branch stars impacts the normalization of the first exponential term in Eq. \ref{eqn:lum_func}.
Recall that the slope $B$ of the exponential term is used to model the population of giant branch stars, and thus as the fainter stars are lost, the exponential flattens to avoid overpredicting the number of these stars.
Then, to match the number of Red Clump, RGBB, AGBB, and brighter giant branch stars, the Gaussian terms that describe the contribution from the Red Clump and RGBB must compensate, resulting in the fit preferring large value of $EW_{RC}$ that push towards our upper limit on the prior of this parameter and deviate $EW_{RC}$ away from a Gaussian distribution.
In addition, the inferred value of $K_{S,RC}$ also shifts to brighter magnitudes in order to better match the bright giant branch stars and lack of faint Red Clump stars.
Essentially, the Gaussian Red Clump term, and to a lesser extent, the Gaussian RGBB and AGBB terms, are attempting to fit the entire population of stars.
An example line of sight where these effects can be seen is shown in Figure~\ref{fig:fits}.

For lines of sight with high extinction $A_{K_S}$ where these effects are important, the fact that the inferred value of $K_{S,RC}$ is biased toward brighter magnitudes implies that our inferred extinction values are underestimated.
We therefore choose to report the values of $A_{K_S}$ inferred from our procedure that suffer from incompleteness as lower limits.  

In order to determine the lines of sight where we only derive lower limits on $A_{K_S}$, we map the distribution of $B$.
In general, we expect $B$ to be roughly constant for a complete sample of giant stars and be well characterized by a Gaussian distribution about its median value $\tilde{B}$.
We determine $\tilde{B} = 0.40$ and that its distribution begins to deviate from Gaussian-like at $B \la 0.2$.
Figure~\ref{fig:bmap} shows the spatial distribution of $B$ and we find that these deviations are mostly restricted to $|b|\leq 0.6^{\circ}$ and $|l|\leq1.2^{\circ}$, indicating that the incompleteness of giant stars is significant in this region.
We flag pixels in our map that have $B<0.2$ and report values of $A_{K_S}$ in these regions as lower limits to the true extinction.

\subsection{Building the Maps} \label{sec:map}

To begin building our maps, we define the resolution for which to create our final maps.
This resolution determines the edge length of our pixels and thus the number of stars included in each color-magnitude diagram.
We want the pixel size to maximize the number of stars available for use in our models.
At the same time, we also want to choose a small enough edge length to retrieve useful location-dependent information and structure from the map.
We settled on a $2^\prime \times 2^\prime$ pixel size, as this resolution appeared to trace the structure of the interstellar dust while also providing a sufficient number of stars necessary to fit.

All stars within an individual pixel are used in the processes outlined in Section \ref{sec:fit}.
We obtain the observed $K_S$-band magnitude and $(H-K_S)$ color of the Red Clump for a given line of sight from these fits.
We compare our observed magnitudes and colors to those of a reference value to determine the extinction and reddening in the given field, a method utilized by many studies \citep[e.g.,][]{1996ApJ...460L..37S,gonz2012, gonz2018, nataf2013, surot2020}.

\begin{figure*}
    \centering
    \includegraphics[width=\linewidth]{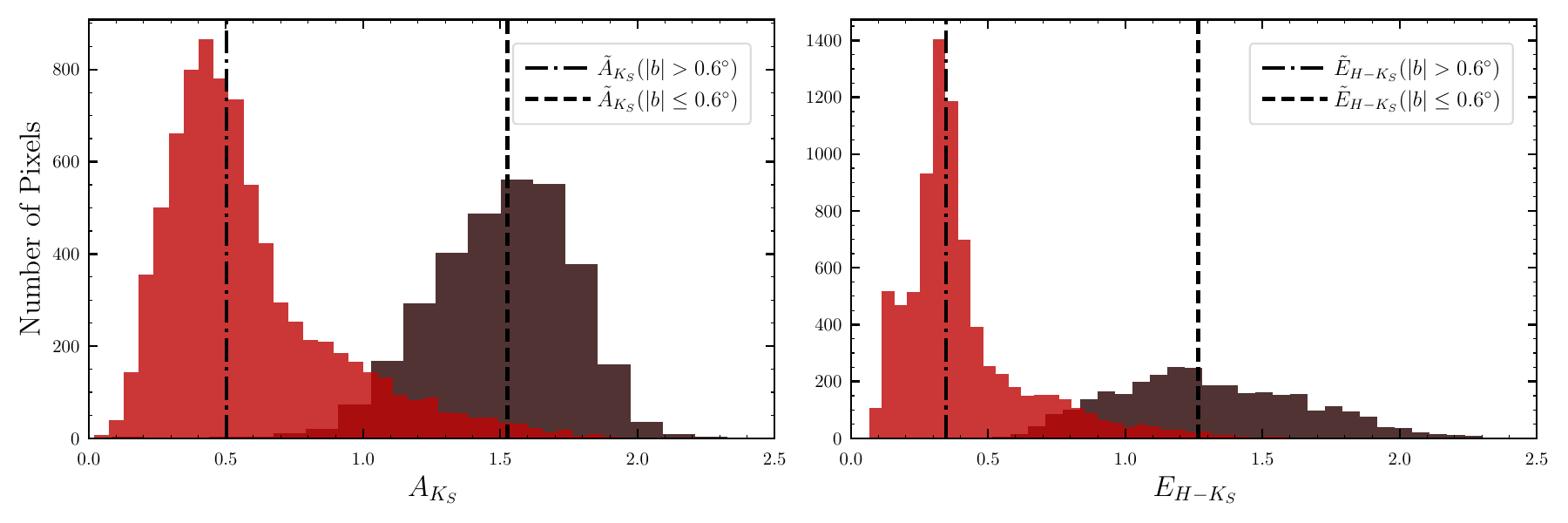}
    \caption{Distributions of the $A_{K_S}$ extinction and $E(H-K_S)$ reddening. Histograms for high $(|b|>0.6^\circ)$ and low $(|b|\leq0.6^\circ)$ latitude fields are shown in red and brown, respectively, to highlight the location dependence of extinction and reddening.}
    \label{fig:akehkshist}
\end{figure*}

To determine the intrinsic magnitude $K_{S,RC}$ and color $(H-K_S)_{RC}$ of the Red Clump, we adopt an absolute magnitude of $M_{K_{S, RC}}=-1.63 \pm0.01$ \citep{2017MNRAS.471..722H, 2021ApJ...910..121N, wang2021} and a distance to the GC of $R_0=8277 \pm 9$ pc \citep{reid2019, 2019A&A...625L..10G, grav2021, grav2022}.

We assume a homogeneous Red Clump population that has a fixed absolute magnitude and distance that depends only on Galactic latitude due to the presence of the bar.
Below $|b|<3^{\circ}$, Red Clump stars are concentrated along the line of sight near the distance to the GC, with their density along the line of sight dropping off rapidly with distance from the GC \citep{wegg2013, wang2021}.
There is a gradient in the distance to the peak density of RC stars along the line of sight with Galactic latitude due to the bar \citep[e.g.,][]{nataf2013, wegg2013, wang2021}, which we account for explicitly below.
This effect is generally small ($\pm ~7\%$) for the UKIRT fields, which probe latitudes between $-2.15 ^\circ \leq l \leq 2.71 ^\circ$.
Variations in the distance distribution of Red Clump stars become significant due to the X-shaped bulge past $|b|>5^{\circ}$ \citep{wegg2013}, which is outside of the UKIRT fields.

The absolute magnitude of Red Clump stars has also been shown to vary with age and metallicity \citep{onozato2019}.
These effects are expected to be minimal in the NIR \citep{salaris2002}, with variations in $M_{K_{S,RC}}$ being observed to a few hundredth magnitudes per Gyr and dex \citep{huang2020, wang2021}.
Outside of $|b|>4^{\circ}$ ($z>600$ pc), a vertical metallicity gradient for Red Clump stars of 0.04 dex/deg has been observed \citep{gonzalez2013}.
The inner 600 pc of the Galactic bulge, however, has been shown to be rather homogenous in metallicity, with variations only within the inner 10 pc \citep{schultheis2019}.
These variations indicate a weak magnitude variation ($\lesssim 0.1$ mag) of clump stars due to metallicity in the lines of sight we probe.

We account for the bar angle by assuming an angle of $\alpha=27\pm2^{\circ}$ \citep{wegg2013, wegg2015}. 
We apply a symmetric geometric correction $(\Delta R_0)$ about $l=0^{\circ}$ to $R_0$ for any given $l$ \citep[see Figure 9 of][]{wegg2013} and adopt the following for $K_{S,RC}$:
\begin{eqnarray}
    \Delta R_0 &=& - \left[ \frac{2\pi R_0}{360^{\circ} \tan \alpha}\right] l  \\
    K_{S,RC} &=& \mu + M_{K_{S, RC}} \nonumber \\
    &=& 5\log_{10}(R_0 + \Delta R_0) - 5 + M_{K_{S, RC}} 
\end{eqnarray}
At $l=0^{\circ}$, this results in a magnitude of $K_{S,RC}=12.96\pm0.01$.
We also adopt a color for the Red Clump of $(H-K_S)_{RC}=0.15 \pm0.01$ \citep{2017MNRAS.471..722H, 2021ApJ...910..121N}, independent of bar angle.

With values for the color and magnitude of the Red Clump, the extinction and reddening are thus given by
\begin{eqnarray}
\centering
A_{K_S} &=& K_{S,RC,obs} - K_{S,RC} \label{eqn:extinction}\\
E(H-K_S)&=&(H-K_S)_{RC,obs}-(H-K_S)_{RC} \label{eqn:colorexcess}
\end{eqnarray}
We apply this to the retrieved magnitudes and color of every line of sight in our map.

The $A_{K_S}$ extinction map of the bulge is shown in Figure~\ref{fig:finalAK}, containing 11,241 pixels each of size $2^\prime \times 2^\prime$.
Figure~\ref{fig:finalEHK} similarly displays the $E(H-K_S)$ reddening of the same fields.
Of these 11,241 pixels, 1,520 pixels have a value of $B<0.2$ that we report as lower limits on the extinction and reddening, with the general region marked in grey, as mentioned in Section \ref{sec:complete}.
We also exclude 9 pixels from our analysis due to the color convergence criterion described in Section~\ref{sec:color}.

Figure~\ref{fig:akehkshist} shows the distributions of the $A_{K_S}$ extinction and $E(H-K_S)$ reddening from our work.
Across all latitudes, the median 1-$\sigma$ uncertainty in $A_{K_S}$ and $E(H-K_S)$ is $\tilde{\sigma}_{A_{K_S}}=0.04$ mag and $\tilde{\sigma}_{E(H-K_S)}=0.06$ mag, respectively.
We find that the median extinction in low latitude fields $(|b|\leq0.6^\circ)$ is $\tilde{A}_{K_S}=1.5^{+0.2}_{-0.3}$ mag and $\tilde{A}_{K_S}=0.5^{+0.4}_{-0.2}$ mag for high latitude fields $(|b|>0.6^\circ)$.
We similarly find the median reddening to be $\tilde{E}(H-K_S)=1.3^{+0.4}_{-0.3}$ mag and $\tilde{E}(H-K_S)=0.3^{+0.2}_{-0.1}$ mag for low and high latitude fields, respectively.
Table~\ref{tab:csv} shows six example lines from the extinction and reddening map we provide along with this publication.

\begin{table*}[]
\caption{Example lines of sight from the UKIRT extinction and reddening maps we provide. $l$ and $b$ are the Galactic longitude and latitude in degrees. $A_{K_S}$ is the $K$-band extinction, and $\sigma_{A_{K_S}}^{+/-}$ provides the upper and lower uncertainties on the value in magnitudes. $E(H-K_S)$ and $\sigma_{E(H-K_S)}$ are the reddening and the square root of the variance in reddening in magnitudes, respectively. ``Flag'' is a boolean that determines whether the field is flagged as a lower limit on the extinction and reddening (Flag=1 when $B<0.2$). \label{tab:csv}}
\begin{tabular}{@{}rrrrrrrr@{}}
\toprule
$l$ (deg) & $b$ (deg) & $A_{K_S}$ & $\sigma_{A_{K_S}}^+$ & $\sigma_{A_{K_S}}^-$ & $E(H-K_S)$ & $\sigma_{E(H-K_S)}$ & Flag \\ \midrule
0.701744  & -1.563869 & 0.341124        & 0.046168                   & 0.044779                   & 0.135057         & 0.034584                  & 0    \\
-0.176280 & 0.220203  & 1.548943        & 0.022379                   & 0.023039                   & 1.172711         & 0.145971                  & 1    \\
0.762704  & -1.865486 & 0.322978        & 0.043215                   & 0.043282                   & 0.157669         & 0.035973                  & 0    \\
0.197118  & -0.780090 & 0.895706        & 0.031732                   & 0.028329                   & 0.771144         & 0.192939                  & 0    \\
1.947216  & 1.348696  & 0.529089        & 0.048652                   & 0.047892                   & 0.364592         & 0.054580                  & 0    \\
0.636030  & -2.643460 & 0.177620        & 0.060590                   & 0.055841                   & 0.125475         & 0.032979                  & 0   \\ \bottomrule
\end{tabular}
\end{table*}

\section{Results} \label{sec:result}

\subsection{Use of extinction laws} \label{sec:law}

\begin{figure}
    \centering
    \includegraphics[width=\linewidth]{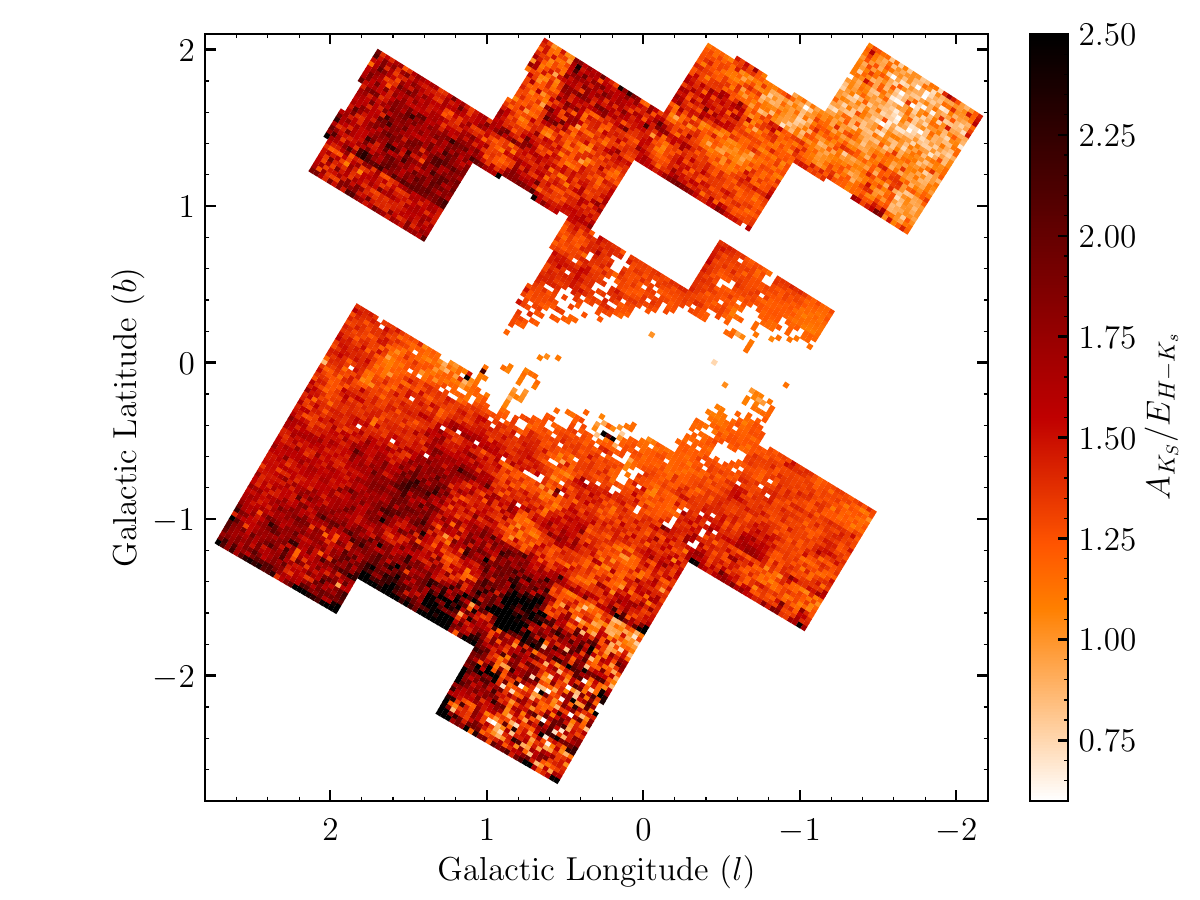}
    \caption{Map of total-to-selective extinction $A_{K_S}/E(H-K_S)$ for each field in the map. We omit fields where $B<0.2$ to highlight the variance in better-characterized fields.}
    \label{fig:rvmap}
\end{figure}

\begin{figure*}
    \centering
    \includegraphics[width=\linewidth]{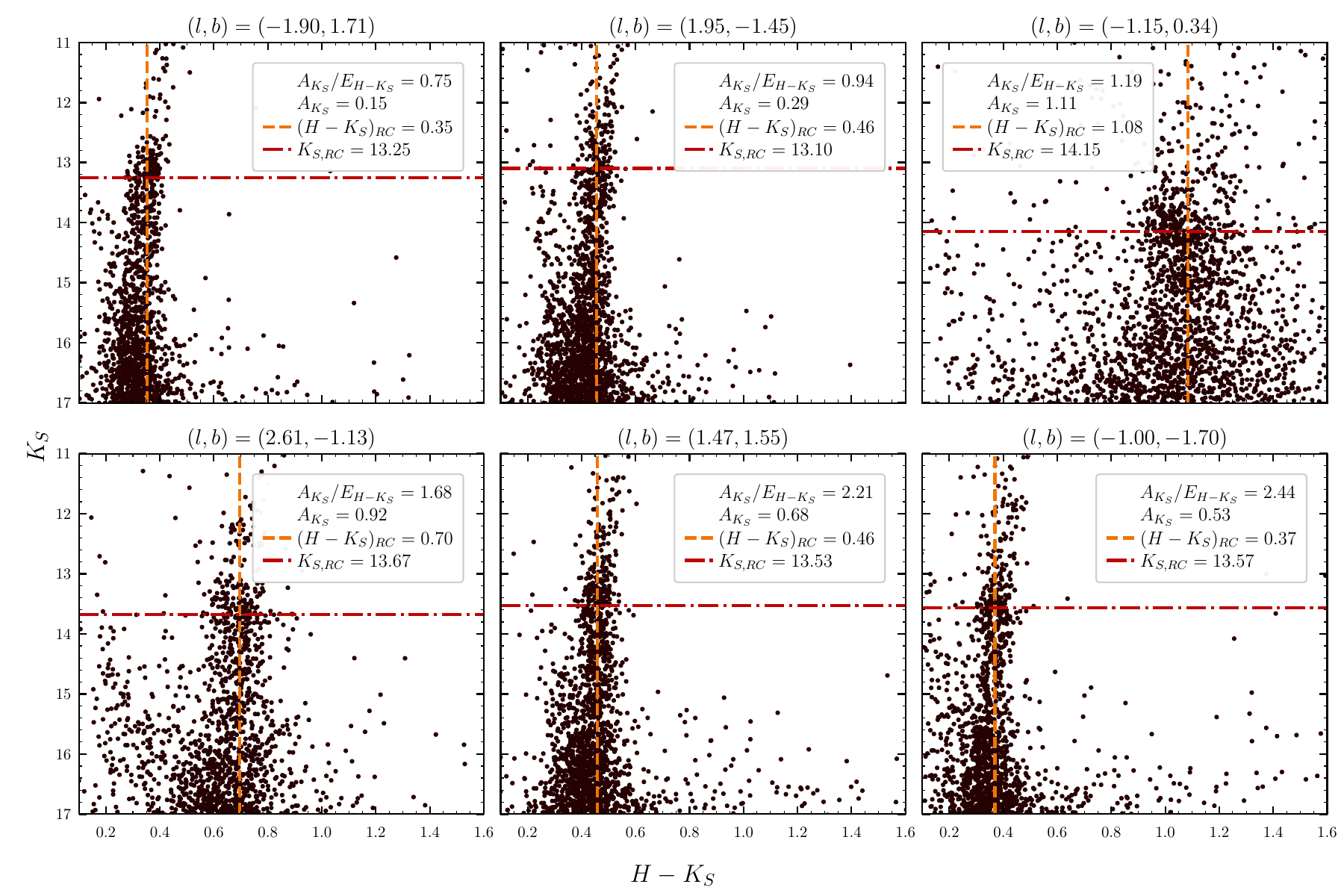}
    \caption{Color-magnitude diagrams of six fields with varying $A_{K_S}/E(H-K_S)$. The color-magnitude diagrams are ordered from smallest $A_{K_S}/E(H-K_S)$ to largest. Each color-magnitude diagram here has a well-populated luminosity function ($B>0.4$). The red dot-dashed and orange dashed lines are the values of $K_{S,RC}$ and $(H-K_{S})_{RC}$ for the given color-magnitude diagram, respectively.}
    \label{fig:6cmd}
\end{figure*}

\begin{figure}
    \centering
    \includegraphics[width=\linewidth]{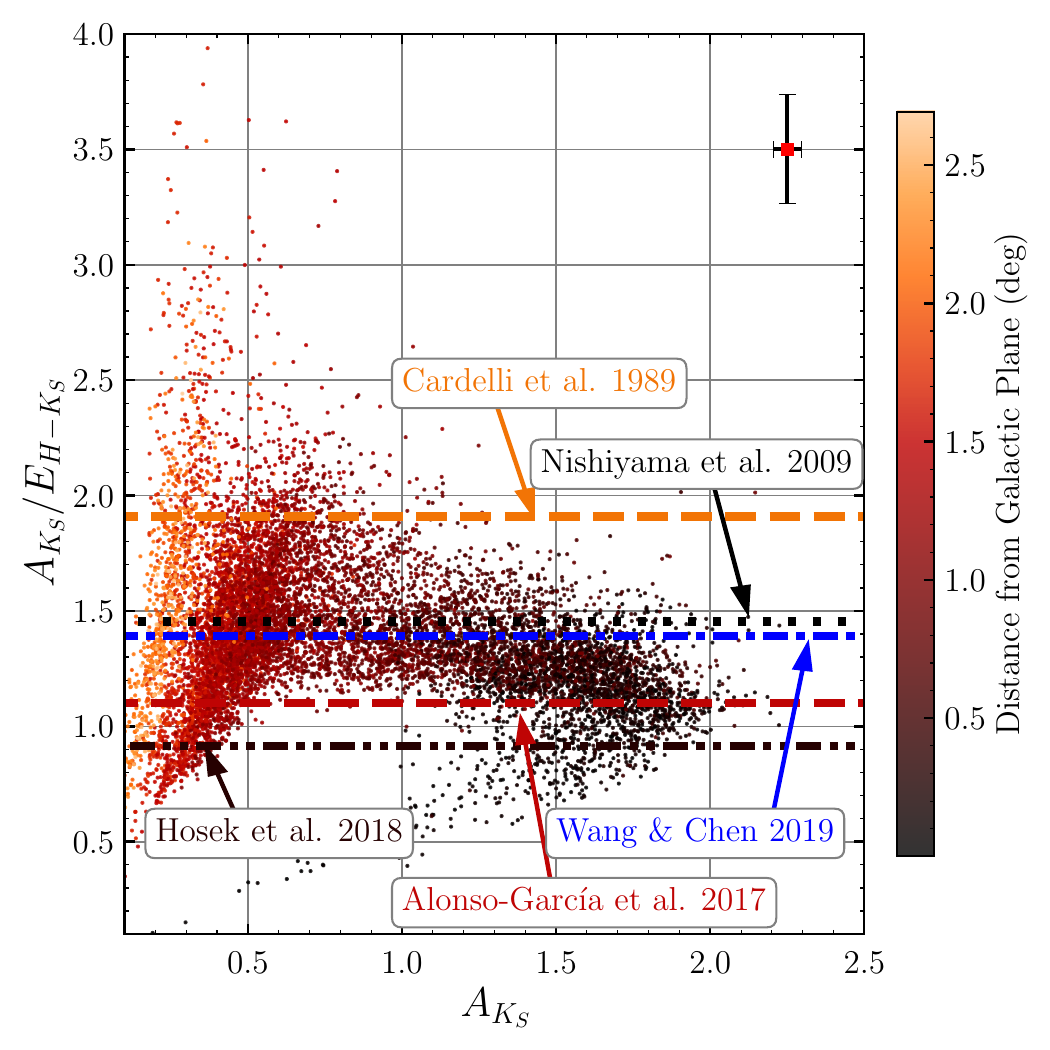}
    \caption{Comparison between various extinction laws against the UKIRT extinction map. Plotted against the $K_S$-band extinction is the ratio of total-to-selective extinction $A_{K_S}/E(H-K_S)$. Points are colored by their distance from the Galactic Plane ($|b|$). The red square in the top right corner shows the median uncertainty in $A_{K_S}/E(H-K_S)$ and $A_{K_S}$.}
    \label{fig:laws}
\end{figure}

Our maps are unique in comparison to many other $K_S$-band extinction maps due to the direct calculation of extinction.
This work does not report extinction values derived using a fixed, standard extinction law (ratio of total to selective extinction or $A_{K_S}/E(H-K_S)$).
Many works use a universal extinction law, with \citet{ccm} being a common standard, to convert from color excess to extinction.
Much work has been conducted to show that the extinction law can vary with location in the Galactic plane \citep[][]{nishiyama2006, nishiyama2009, schlafly2016}.
We find and corroborate significant variation in the extinction law with Galactic latitude \citep{nishiyama2009}.

By independently determining the extinction and color excess at each location, we can also investigate the validity of applying universal extinction laws and the variation of the extinction law with location.
We compare extinction values found from our luminosity function fit (see Sections \ref{sec:lum}-\ref{sec:mcmc}) to those from our color excess (see Section \ref{sec:color}) using various extinction laws.
Figure~\ref{fig:rvmap} shows a map of the total-to-selective extinction, $A_{K_S}/E(H-K_S)$.
For this map, we have excluded fields with $B < 0.2$, for which our estimate of the extinction is likely unreliable.
Taking all lines of sight into account, we report a median value of $A_{K_S}/E(H-K_S)$ to be $1.4^{+0.3}_{-0.2}$.
We also find a median 1-$\sigma$ uncertainty of $\tilde{\sigma}_{A_{K_S}/E(H-K_S)} = 0.2$ in our $A_{K_S}/E(H-K_S)$ measurements.

Figure~\ref{fig:6cmd} displays the color-magnitude diagrams of six of these fields, each with varying $A_{K_S}/E(H-K_S)$.
Each fit of these color-magnitude diagrams resulted in an exponential parameter $B>0.4$, ensuring that the Red Clump is well determined.
Figure~\ref{fig:6cmd} shows how different lines of sight result in unique stellar populations that cannot be described using a single treatment.

We analyze and compare the NIR extinction laws by \citet{ccm}, \citet{nishiyama2009}, \citet{alonsogarcia2017}, \citet{hosek2018}, and  \citet{wangchen2019}, which report or derive $A_{K_S}/E(H-K_S)$ values of 1.91, 1.46, 1.10, 0.91, and 1.39, respectively.
Appendix \ref{app:ext} gives a table of the laws derived from each of these works for multiple filter sets.
The extinction law of \citet{wangchen2019}, $A_{K_S}/E(H-K_S)=1.39$, best matches the median $A_{K_S}/E(H-K_S)$ we report of $1.4^{+0.3}_{-0.2}$.
Figure~\ref{fig:laws} shows the total-to-selective extinction ratio to the $K_S$-band extinction for every pixel, colored by latitudinal displacement from the Galactic plane.
The red square in the top right of the figure shows the median 1-$\sigma$ uncertainty in $A_{K_S}/E(H-K_S)$ and $A_{K_S}$.
Shown as horizontal lines are the total-to-selective extinction ratio, $A_{K_S}/E(H-K_S)$, for each of the works.
We find that a universal value of the ratio of total to selective extinction does not describe our data well, corroborating the findings of \citet{nishiyama2006, nishiyama2009}.
For example, adopting a universal extinction law at fixed $E(H-K_S)$ leads to an overestimate of the extinction for high $|b|$ and an underestimate of the extinction at low $|b|$.

\subsection{Comparison to literature}\label{sec:compare}

\begin{figure*}
    \centering
    \includegraphics[width=\linewidth]{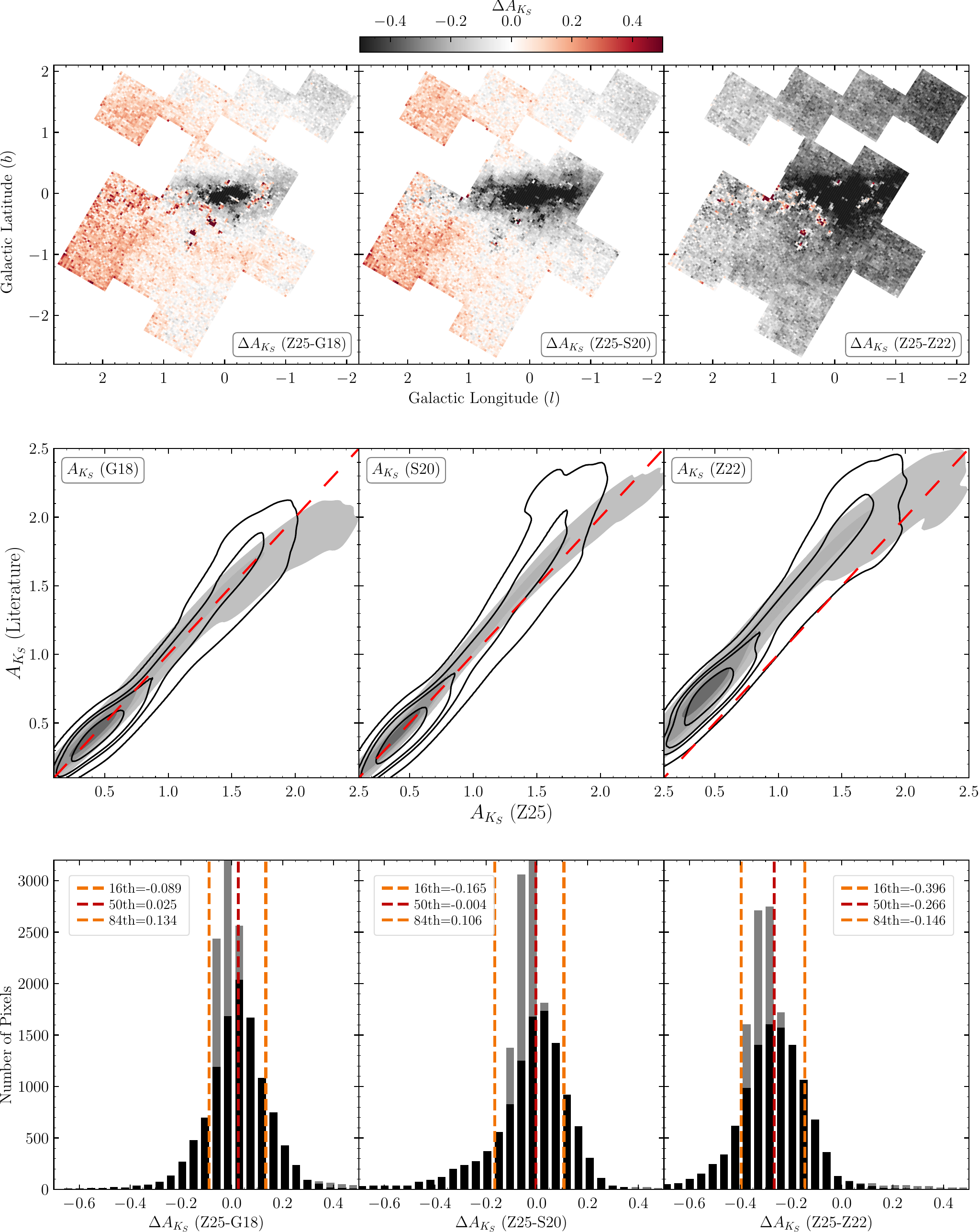}
    \caption{Comparison of extinction maps from this work with those from G18, S20, and Z22. The top row shows the difference between our extinction map, derived from fitting Equation~\ref{eqn:lum_func}, and the literature maps, within the UKIRT fields. The middle row displays extinction distributions using a Gaussian kernel density estimator: black open contours represent values from the luminosity function fit, while gray filled contours show extinction derived from color excess and the \citet{wangchen2019} extinction law. The bottom row shows histograms of the differences in extinction values, with black bars for the luminosity function fit and gray bars for the color excess method. Dashed lines indicate the percentiles of $\Delta A_{K_S}$ from the luminosity function-based extinction.}
    \label{fig:compare}
\end{figure*}

The published works of \citet{schultheis99, gonz2012, wegg2013, gonz2018, surot2020, zhang2022} are good benchmarks for the NIR extinction and reddening towards the bulge.
We selectively compare our extinction and reddening maps (hereafter Z25) against the most recent of these works, namely \citet{gonz2018, surot2020, zhang2022} (G18, S20, and Z22, respectively).
These maps all report inferred extinction using extinction laws or suggest using one to convert from color excess to extinction.
All the works above primarily use data from the Vista Variables in the Via Lactea (VVV) survey, thus making our work a useful cross-check as well.

To compare the maps of the previously published works to ours, we perform a pixel-by-pixel linear interpolation to map their reddening and extinction values onto the grid centers used in our UKIRT maps.
For consistency, we use the same extinction law to compare the reddening in their works to the extinction in our own.
Z22 utilizes the extinction law from \citet{wangchen2019} to obtain their extinction values and provide those instead of color excess.
Thus, we also convert G18 and S20 using this law to limit any variation that would arise from the difference in laws used.
We would like to note the bias that using a universal extinction law causes, as described in Section \ref{sec:law}.

{We compare two different methods for measuring the $A_{K_S}$ extinction using our analysis in the UKIRT fields to the values determined by G18, S20, and Z22.
First, we compare the $A_{K_S}$ values we derived by directly fitting the $K_S$ magnitude distribution of clump stars to the model luminosity function (Equation~\ref{eqn:lum_func}).
This method allows for variable values of $A_{K_S}/E(H-K_S)$ in our analysis.
Second, we multiply our measured values of the $E(H-K_S)$ color excess in each field by the \citet{wangchen2019} value of $A_{K_S}/E(H-K_S)=1.39$, thereby assuming a constant extinction law.
Figure~\ref{fig:compare} displays the difference between the $K_S$-band extinction found in our work and the works of G18, S20, and Z22 after applying the $A_{K_S}/E(J-K_S)=0.49$ extinction law from \citet{wangchen2019}.
We find overall reasonable agreement with G18 and S20, but find a global offset in our inferred values of $A_{K_S}$ relative to Z22.
We find that our maps consistently report lower values of $A_{K_S}$ towards 0 degrees in $l$ and $b$.

Figure~\ref{fig:rvcomp} displays how variations in the measured $A_{K_S}/E(H-K_S)$ compare to the magnitude differences between each survey.
The top row displays extinction values from directly fitting for the $K$-band magnitude from Equation~\ref{eqn:lum_func}, while the bottom row is the magnitude difference from applying \citet{wangchen2019} to $E(H-K_S)$ to convert from color excess to extinction.
In the top left corner of the left-most panel in each row is a red square that shows the median uncertainty in $A_{K_S}/E(H-K_S)$ and $A_{K_S}$ for the respective method.
The median uncertainty in $A_{K_S}$ does not account for uncertainties in the maps being compared against, but does account for uncertainty from the \citet{wangchen2019} extinction law in the bottom row.

If the conversion from color excess to single-band extinction were universal for all fields, there would be no vertical dispersion in $A_{K_S}/E(H-K_S)$.
At high latitudes, the tight spread about $\Delta A_{K_S}=0$ between the color excess-derived extinction in the bottom row highlights the consistency of our derived $E(H-K_S)$ with the literature.
The vertical dispersion of $A_{K_S}/E(H-K_S)$ for these high latitude fields shows a strong dependence on $A_{K_S}$, highlighting the presence of a variable extinction law.

In low latitude fields, we consistently report higher values of $E(H-K_S)$ and lower $A_{K_S}$ than the literature.
We find that using a universal extinction law in these low latitude fields will consistently overestimate the single-band extinction.
We also note that this trend at low latitudes may be mildly exaggerated by underestimates of $A_{K_S}$ caused by a lack of faint bulge stars in these fields.
Nonetheless, the vertical spread of $A_{K_S}/E(H-K_S)$ across all latitudes clearly highlights the presence of a variable extinction law and the inability of universal extinction laws to describe the relationship between extinction and reddening.
The use of other universal extinction laws would only shift the center of $\Delta A_{K_S}$ and not change the spread of magnitude differences.
Below, we describe the specifics for each of the extinction maps we compare against individually.

\subsubsection{\citet{gonz2012, gonz2018}} \label{sec:gonz}

The $E(J-K_S)$ maps derived by \citet{gonz2012, gonz2018} (G12 and G18) were created by similarly using Red Clump stars in Baade's Window as standard crayons \citep{2011A&A...534A...3G}.
Both works use the same method outlined in \citet{2011A&A...534A...3G}, but the VVV catalogs utilized vary.
G12 used photometric data from the 2MASS survey to complete the brighter ends of their data, due to saturation in VVV \citep{2011A&A...534A...3G}.
The maps of G18 use VVV data similar to G12 but with deeper and more complete PSF photometry \citep{gonz2018}.
Thus, we do not compare our work directly to G12 here and compare it to G18.

Our results from fitting Equation~\ref{eqn:lum_func} differed from those of G18 by a median of $\tilde{\Delta A_{K_S}}(Z25-G18)=0.025_{-0.115}^{+0.109}$ mag.
Our color excess-derived extinction differs from G18 by $\tilde{\Delta A_{K_S}}(Z25-G18)=-0.005_{-0.049}^{+0.065}$ mag.
As shown in the top panel of Figure~\ref{fig:compare}, we underestimated the extinction of sight lines at low latitudes where UKIRT suffers from completeness (see Section \ref{sec:complete}).
Towards regions of low extinction, our maps report slightly higher values of $K_S$-band extinction than those of G18.

\subsubsection{\citet{surot2020}} \label{sec:surot}

\citet{surot2020} present high-resolution extinction maps using their own MW-BULGE-PSPHOT catalog \citep{surot2019}.
S20, similarly to G18, uses fits of the $(J-K_S)$ color of the Red Clump along with a reference color from Baade's Window to determine $E(J-K_S)$.
They manually selected regions in the color-magnitude diagram that would maximize the number of Red Clump and red giant branch stars while minimizing foreground contamination.
This region is selected by following the reddening vector for each tile.
Instead of adopting a universal reddening vector, they derive their own following a similar procedure to that of \citet{alonsogarcia2017} for each VVV tile.

For each of their fields, they find the 20 nearest stars and weight their color contributions to the pixel average based on the distance to the pixel center.
They then use this average color to report the color excess and calibrate it against G18.
S20 reports $E(J-K_S)$ rather than the extinction.
Without the reddening vector they derived for each grid provided, which would have helped limit the bias resulting from a constant conversion, we therefore simply apply \citet{wangchen2019} to obtain a $K_S$ extinction.
This choice also helps limit variations in the comparison between their work with G18 and Z22.

With S20 sharing many of the same characteristics as G18, it isn't a surprise that our results have similarities when compared.
We find a median difference in the Equation~\ref{eqn:lum_func} derived extinction between our maps of $\tilde{\Delta A_{K_S}}(Z25-S20)=-0.004^{+0.110}_{-0.161}$ mag.
We also find a median difference in the color excess-derived extinction of $\tilde{\Delta A_{K_S}}(Z25-S20)=-0.029^{+0.050}_{-0.055}$ mag.
In regions where UKIRT suffers from completeness (see Section \ref{sec:complete}), we consistently report lower values of extinction compared to S20.
Outside of these low-latitude regions, we trace equivalent or slightly higher values of extinction.

\begin{figure*}[t]
    \centering
    \includegraphics[width=\linewidth]{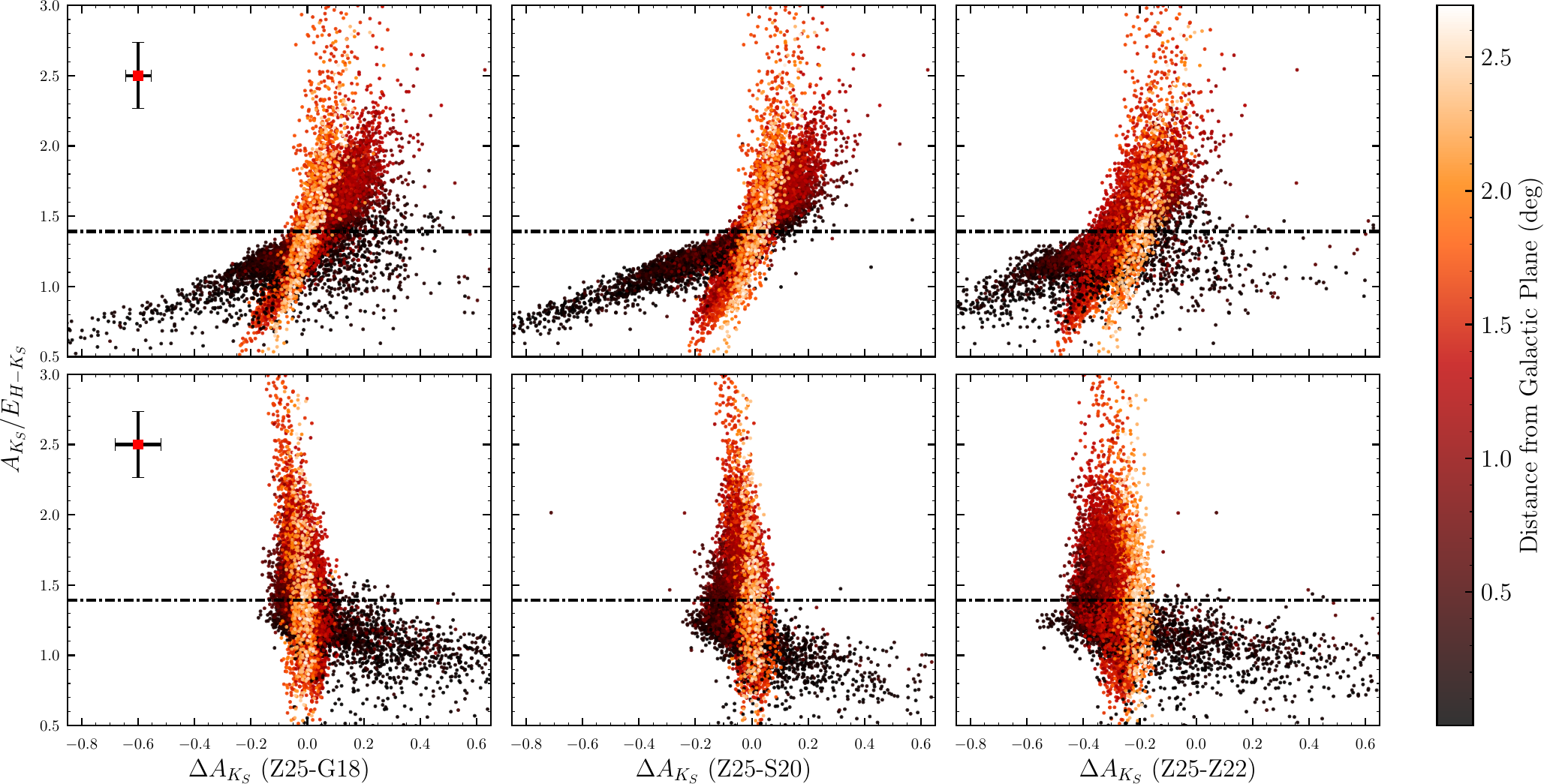}
    \caption{Comparison of $A_{K_S}/E(H-K_S)$ from our work to the difference from our extinction values with \citet{gonz2018}, \citet{surot2020}, and \citet{zhang2022}. Each point is colored by vertical distance from the Galactic Center in degrees. The top row shows $\Delta A_{K_S}$ from the luminosity function fit-derived extinction, while the bottom panel used \citet{wangchen2019} to convert color excess into extinction. The red square in the top left corner shows the median uncertainty on $A_{K_S}/E(H-K_S)$ and $A_{K_S}$ from both methods. The median uncertainty in $A_{K_S}$ does not account for contributions from the literature being compared against, but does account for the uncertainty in the \citet{wangchen2019} extinction law.}
    \label{fig:rvcomp}
\end{figure*}

\subsubsection{\citet{zhang2022}}

\citet{zhang2022} approached their extinction estimates using a method they call XPNICER, a combination of an unsupervised machine learning technique coined PNICER \citep{meingast2017} and the color estimation method X percentile \citep{dobashi2008}.
They conducted this work using their own VVV catalogs created using the DaoPHOT algorithm \citep{stetson1987}.
These photometric catalogs were initially calibrated using the prior VVV catalogs of \citet{alonsogarcia2018} but were recalibrated according to the suggestions of \citet{hajdu2020} to account for photometric zero-point biases.
They assume a universal extinction law across the entire VVV field, adopting the value from \citet{wangchen2019}.
For a complete overview of their methodology, see Section 3 of Z22.

Our luminosity function-derived extinction differs from that of Z22 by a median value of $\tilde{\Delta A_{K_S}}(Z25-Z22)=-0.266^{+0.120}_{-0.129}$ mag.
We also find a median difference in the color excess-derived extinction of $\tilde{\Delta A_{K_S}}(Z25-Z22)=-0.297^{+0.080}_{-0.065}$ mag.
Z22 noticed a similar difference between their work and G18 and S20, with a translational offset in magnitude space.
The right panel of the middle row of Figure~\ref{fig:compare} clearly shows this translational offset.
Z22 mentioned that a difference in the photometric zero-points of G18 and S20 or stellar populations used could be the result of these discrepancies.
Since our work also shows a similar translational offset while using a different survey, the difference in the zero-point is likely not the most dominant factor.
This discrepancy in our works likely originates from the difference in stellar populations used between Z22 and the works of G18, S20, and our own.
Z22 pointed this out themselves; their study includes the foreground main-sequence stars, while G18, S20, and our work utilize just the Red Clump.
Z22 also noted that their methodology overestimated the $A_V$ source extinction by roughly 1 mag, which may contribute to the translational offset observed.

\section{Conclusion}

We present high-resolution $2^\prime \times 2^\prime$ maps of the Galactic Bulge, spanning from $l=-2.15^{\circ}$ to $2.71^{\circ}$ and $b=-2.69^{\circ}$ to $2.03^{\circ}$, using dual $H$ and $K_S$-band photometry from the 2017 UKIRT microlensing survey.
We compared the measured value of the $K_S$-band magnitude and $H-K_S$ color of the Red Clump for each pixel to the intrinsic magnitude and color to infer $A_{K_S}$ and $E(H-K_S)$ for each line of sight.
In regions towards the Galactic Plane (between $|b|\leq 0.6^{\circ}$ and $|l|\leq1.2^{\circ}$) we report lower limits on the extinction due to the completeness issues with the UKIRT microlensing survey in these highly extinguished fields.

Our maps show good agreement with the VVV maps of \citet{gonz2012, gonz2018} as well as \citet{surot2020}.
We also observe the same offset in $A_{K_S}$ with the work by \citet{zhang2022} that is consistent with their comparison with \citet{gonz2012, gonz2018} and \citet{surot2020}.
Regions in which we find the most disagreement are those towards the Galactic Center, where the UKIRT data suffers most from completeness.
The use of universal extinction laws to convert from color excess to extinction was also explored.
We confirm previous findings that toward the Galactic bulge the ratio of total-to-selective extinction $A_{K_S}/E(H-K_S)$ is not universal, but rather depends on the line of sight, varying strongly with location in the Galactic Plane.

\begin{acknowledgments}

We would like to thank the anonymous referee for suggestions that significantly improved the quality of the paper.

UKIRT is currently owned by the University of Hawaii (UH) and operated by the UH Institute for Astronomy; operations are enabled through the cooperation of the East Asian Observatory. When some of the 2017 data reported here were acquired, UKIRT was supported by NASA and operated under an agreement among the University of Hawaii, the University of Arizona, and Lockheed Martin Advanced Technology Center; operations were enabled through the cooperation of the East Asian Observatory. The collection of the 2017 data reported here was furthermore partially supported by NASA grants NNX17AD73G and NNG16PJ32C \citep{NEAMicrolensing}.

This paper makes use of data from the UKIRT microlensing surveys \citep{2017AJ....153...61S} provided by the UKIRT Microlensing Team and services at the NASA Exoplanet Archive, which is operated by the California Institute of Technology, under contract with the National Aeronautics and Space Administration under the Exoplanet Exploration Program.

The authors wish to recognize and acknowledge the very significant cultural role and reverence that the summit of Mauna Kea has always had within the indigenous Hawaiian community. We are most fortunate to have the opportunity to use data produced from observations conducted on this mountain.

This research has made use of the SVO Filter Profile Service ``Carlos Rodrigo'', funded by MCIN/AEI/10.13039/501100011033/ through grant PID2023-146210NB-I00.

B.S.G. acknowledges the support of the Thomas Jefferson Chair for Discovery and Space Exploration at the Ohio State University.
D.M.N. acknowledges support from NASA under award Number 80NSSC21K1570 and award Number 80NSSC19K058.
A.S.Z., S.A.J., and B.S.G. were partially supported by NASA grant NNG16PJ32.

\end{acknowledgments}

\begin{contribution}

A.S.Z. led the analysis, developed the software, and wrote the initial draft of the manuscript. They also prepared figures and handled the submission process.

S.A.J. advised A.S.Z., contributed to the initial analysis framework and manuscript draft, and provided detailed feedback on the manuscript.

B.S.G. advised both A.S.Z. and S.A.J., verified key aspects of the analysis, contributed to significant manuscript revisions, and provided funding support for the project.

G.B. supervised the early stages of the project before A.S.Z.’s involvement and contributed observational data.

D.M.N. contributed to the creation and validation of the analysis techniques and provided comments on the manuscript.

Y.S. contributed to calibrating the observational data.

\end{contribution}

\section*{Data Availability}

Our extinction and reddening maps are available online as a CSV file on Zenodo via \url{https://doi.org/10.5281/zenodo.14919461}. Table~\ref{tab:csv} shows six example lines of the map we provide.

\appendix

\section{Extinction Laws}\label{app:ext}

\begin{table*}
\caption{Total-to-selective extinction ratios for each of the extinction law papers we use to investigate. We also provide calculations for the proposed filters used in the upcoming \textit{Roman} Galactic Bulge Time Domain Survey as well as the filters used by the recently launched \textit{Euclid}.}
\label{tab:law}
\centering
\resizebox{\linewidth}{!}{%
\begin{tabular}{@{}lccccccccc@{}}
\toprule
\multicolumn{1}{c}{}     & \multicolumn{2}{c}{UKIRT} & \multicolumn{2}{c}{2MASS} & \multicolumn{2}{c}{VVV} & \textit{Roman} & \multicolumn{2}{c}{\textit{EUCLID}} \\ \cmidrule{2-10}
                         & $A_K/E(J-K)$       & $A_K/E(H-K)$       & $A_{K_S}/E(J-K_S)$     & $A_{K_S}/E(H-K_S)$     & $A_{K_S}/E(J-K_S)$    & $A_{K_S}/E(H-K_S)$    & $A_{F146}/E(F087-F146)$    & $A_H/E(Y-H)$            & $A_H/E(J-H)$            \\
\citet{ccm}              & 0.670       & 1.614       & 0.686       & 1.909       & 0.720      & 1.828      & 1.052          & 0.838            & 1.922            \\
\citet{nishiyama2009}    & 0.474       & 1.219       & 0.486       & 1.455       & 0.513      & 1.390      & 0.784          & 0.605            & 1.465            \\
\citet{alonsogarcia2017} & 0.327       & 0.913       & 0.336       & 1.101       & 0.357      & 1.049      & 0.568          & 0.428            & 1.109            \\
\citet{hosek2018}        & ---         & ---         & 0.347       & 0.913       & 0.378      & 0.971      & ---            & ---              & ---              \\
\citet{wangchen2019}     & 0.447       & 1.164       & 0.459       & 1.391       & 0.485      & 1.329      & 0.745          & 0.573            & 1.402            \\ \bottomrule
\end{tabular}}%
\end{table*}

Here we place our derivations of the NIR extinction laws used in our work.
We calculate and provide laws from various literature for specific telescope filters used in this work, as well as some others for reference, given in Table \ref{tab:law}.
Effective wavelengths for each filter were collated from the Spanish Visual Observatory's Filter Profile Service \citep{2020sea..confE.182R}.
For \textit{Roman} we chose to only include the proposed filters for the Galactic Bulge Time Domain Survey, with \textit{F146} being the primary observing filter and \textit{F087} proposed for a lower cadence \citep{2019ApJS..241....3P, wilson2023}.
For the works of \citet{ccm, nishiyama2009, alonsogarcia2017, wangchen2019} we derive the extinction laws from either equations they provide or by using their derived power law.
\citet{hosek2018} does not provide either, so we just provide the VVV and 2MASS values from their work here.

\bibliography{sample7}{}
\bibliographystyle{aasjournalv7}

\end{document}